\documentclass[12pt,preprint]{aastex}

\slugcomment{AJ, accepted Dec. 10, 2003}

\shorttitle{The Ionized Gas in Starbursts}
\shortauthors{Calzetti et al.}

\begin{document}

\title{The Ionized Gas in Local Starburst Galaxies:\\ 
Global and Small--Scale Feedback from Star Formation.\altaffilmark{1}}

\author{Daniela Calzetti, Jason Harris}
\affil{Space Telescope Science Institute, Baltimore, MD 21218}
\email{calzetti,jharris@stsci.edu}

\author{John S. Gallagher, III}
\affil{Department of Astronomy, University of Wisconsin, Madison, WI 53706}

\author{Denise A. Smith\altaffilmark{2}}
\affil{Space Telescope Science Institute, Baltimore, MD 21218}

\author{Christopher J. Conselice}
\affil{California Institute of Technology, Pasadena, CA 91125}

\author{Nicole Homeier}
\affil{Department of Astronomy, University of Wisconsin, Madison, WI 53706}

\and

\author{Lisa Kewley}
\affil{Harvard--Smithsonian Center for Astrophysics, Cambridge, MA 02138}

\altaffiltext{1}{Based on observations obtained with the NASA/ESA
Hubble Space Telescope at the Space Telescope Science Institute, which
is operated by the Association of Universities for Research in
Astronomy, Inc., under NASA contract NAS5-26555. }
\altaffiltext{2}{Computer Science Corporation}

\begin{abstract}

The small-- and intermediate--scale structure and the fraction of the
ISM ionized by non--radiative processes is investigated in a small
sample of four local starburst galaxies, imaged with the Hubble Space
Telescope Wide Field and Planetary Camera~2. The sample comprises
three dwarf galaxies, NGC3077, NGC4214, and NGC5253, and one giant
spiral, NGC5236 (M83). The galaxies span a range in metallicity
($\sim$0.2--2~Z$_{\odot}$), luminosity (M$_B \sim -$17 -- $-$20), and
environment (isolated, interacting), enabling the investigation of
non--radiative ionization processes in a variety of galactic
conditions. For this purpose, the four galaxies were imaged in the
lines of H$\beta$(4861~\AA), [OIII](5007~\AA), H$\alpha$(6563~\AA),
and [SII](6717,6731~\AA). This is a unique set of data, as very few
galaxies (and only our four starbursts) have ever been imaged by HST
in the relatively faint lines of H$\beta$ and [SII]. The use of the
HST has allowed us to trace non--photoionized gas in these galaxies on
scales ranging from a few tens of pc to a few hundred pc, and thus
provide a full budget for this ionized gas component. Using the
`maximum starburst line' of \citet{kewl01} to discriminate between
photoionized and non--photoionized gas, we find that in all four
galaxies non--photoionization processes are responsible for a small
fraction of the total H$\alpha$ emission, at the level of 3\%--4\%.
Because the non--photoionized gas is associated with low H$\alpha$
surface brightness, it occupies between 1/6 and 1/4 of the total
imaged area. The central starbursts yield enough mechanical energy to
produce the non--photoioned gas in the four galaxies, via shocks from
massive star winds and supernova explosions. In particular, the
starbursts in the three dwarf galaxies deposit a significant fraction,
70\%--100\%, of their mechanical energy into the surrounding
interstellar medium (ISM), in order to account for the observed
luminosity of the non--photoionized gas. The morphology of the
non--photoionized regions is different in the dwarfs and the giant
spiral. As already established in previous works, non--photoionized
gas in dwarfs is mainly associated with extended `shells' or
filamentary regions, likely areas of supernova--driven expanding
gas. In all three dwarfs star formation has been an ongoing process
for the last few$\times$10$^7$~yr to $\sim$10$^8$~yr; time--extended
star formation episodes are a requirement to sustain the observed
luminosity of the non--photoionized gas. In the massive spiral, the
non--photoionized gas is concentrated in localized areas surrounded by
active star--formation, with no evidence for extended structures on
the same (or smaller) spatial scales as (than) the `shells' in the
dwarfs. The two H$\alpha$ cavities in NGC5236 may be evolved regions
within the starburst. This confirms the picture that starbursts remain
confined events in massive galaxies, likely due to the deep potential
well.
\end{abstract}


\section{Introduction}

One missing piece for placing star formation within the broader
context of galaxy evolution is the quantification of the feedback
mechanisms between galactic--scale star formation and the host
galaxy's interstellar medium (ISM). One of the extant questions is:
What are the regulating mechanisms for the production of structures at
all scales (associations, rings, bubbles, superbubbles)? In a
star--formation event, stellar winds and supernova explosions from
massive stars inject metals and kinetic energy into the
surrounding ISM. The energy injected may produce gas outflows and, in
more extreme cases, superwinds \citep{heck90,ceci02}, which may act as
regulating mechanisms by removing gas from the site of star formation
and quenching the star formation itself \citep{elme92,meur97}. Such
feedback mechanisms may have strong influence on the evolution of
dwarf galaxies \citep{deke86,marco94}, although the exact details of
the process are still uncertain and depend on the nature/geometry of
the galaxy \citep{deyo94,macl99}.  OB associations will drive
ionization and shock fronts through the ISM that may in some cases
cause the star formation to propagate spatially
\citep{elme77,mccr87,puxl97}.

Adding pieces to the puzzle requires exploring the feedback mechanisms
on all scales, from the $\sim$100--1000~pc scales of superbubbles and
outflows/superwinds down to the $\sim$10~pc scales typical of the
interfaces between the massive stars and the ionization and shock
fronts. While the former has received the most attention in the past
years because of its macroscopic effects on the host galaxy's ISM
\citep{heck90,mart98,ceci02}, the latter has been so far only coarsely
explored \citep{mart97,calz99}, despite its key role for understanding
the large scale processes. In this paper, we employ Hubble Space
Telescope high angular--resolution images of nearby starburst galaxies
to investigate the role of non--radiative ionization processes within
and around sites of star formation. We quantify the amount of
non--photoionized gas present in the starburst regions of these
galaxies, investigate the influence of external environmental factors
on the gas properties/structure, and relate the gas morphology to that
of the young stellar populations.

Throughout the paper, we define as `photoionized' the gas ionized by
radiative processes, and as `non--photoionized' the gas ionized by any
other mechanism. There are a number of processes other than radiative
that can produce observable ionized gas in galaxies. Shocks, and their
precursors, from supernovae and massive star winds
\citep{shul79,alle99} are a viable mechanism in the presence of
intense star formation. Alternative mechanisms are turbulent mixing
layers \citep{slav93} and changes in the gas temperature. The latter
has been invoked to explain the observed forbidden line ratios in
galaxy haloes, at scale heights from $\sim$1~kpc to 5--6~kpc above the
galactic plane of the Milky Way and other quiescently star--forming
galaxies \citep{rand98,haff99,otte99,otte02}. The present analysis
will not enable us to unambiguously discriminate between different
non--photoionization processes; we will tend to privilege the
mechanical input from supernovae and massive star winds, thus shocks,
as the simplest mechanism for driving non--photoionization and
producing the observed morphology and scale (a few 100~pc to
$\sim$kpc) of the non--photoionized gas in the starburst galaxies
\citep{mart97,mart98}). A more detailed justification for this choice
is given in section~6.1.

One result of this study is that the non--photoionized gas represents
a small fraction, about 3\%--4\%, and possibly up to
$\approx$10\%--20\% depending on the assumptions, of the integrated
emission line spectrum in the starbursts. Thus, the presence of
non--photoionized gas will have small or negligible effect on the
interpretation of diagnostics of distant galaxies. However, even such
small fraction is key for tracing the location and morphology of
possible large--scale shock structures in spatially resolved
data. These structures indicate whether and where large amounts of
mechanical energy are being deposited in the interstellar medium, and
where larger-scale phenomena (e.g., superwinds) may start.

The sample consists of four starburst galaxies closer than 5~Mpc, that
we imaged in the light of hydrogen and metal forbidden line
emission. The basic characteristics of the four galaxies are listed in
Table~\ref{tbl-1}. Three of the galaxies, NGC3077, NGC4214, and
NGC5253, are dwarf irregulars (M$\sim$10$^{9}$~M$_{\odot}$), with
similar blue magnitudes, although with different metallicities and
infrared--to--blue ratios. NGC4214 is a blue, relatively
low--extinction, low--metallicity, and isolated galaxy. NGC5253 is
likely in interaction with NGC5236 (M83); although NGC5253 is as blue
and as metal--poor as NGC4214, it has about 60\% more infrared
emission than the latter, an indication of its higher dust content or
concentration.  NGC3077 is a member of the M81 group, forming a
triplet together with M81 and M82, and has a metallicity about 5 times
higher than the other two dwarfs in the sample, with a value close to
solar. NGC5236, the galaxy in likely interaction with NGC5253, is also
in our sample, and is a massive (M$\sim$10$^{11}$~M$_{\odot}$),
grand--design, barred spiral, seen almost face--on.

The present paper is organized as follows. Section~2 provides a brief
description of the four galaxies in the sample. Section~3 describes
the observations and the data reduction, with emphasis on the
derivation of emission--line images. Section~4 presents the derivation
of the line ratio images. Section~5 describes how non--photoionized
gas is discriminated from photoionized gas and presents the main
findings of this work. Section~6 discusses the results and section~7
summarizes them.

\section{Sample's Description}

\subsection{NGC~5253}

NGC5253 is a benchmark starburst with centrally concentrated star
formation superimposed on an older, quiescent stellar population. The
galaxy is possibly in interaction with NGC5236, given the small radial
(500~kpc) and projected (130~kpc) distance between the two
\citep{thim03}. A close encounter between the two galaxies about 1~Gyr
ago was first proposed by \citet{vand80} on the basis of various
evidence, including the warping of the HI disk of NGC5236
\citep{rogs74}. NGC5253 is at the same time a strong UV and FIR
emitter because the starburst's spectral energy distribution is very
blue, but is also crossed by dust lanes that produce patchy
obscuration \citep{kinn93,tele93,calz97}. Radio observations indicate
that a large fraction of the most recent star formation is hidden by
dust \citep{turn98,turn00}. The main E--W dust lane is also the
location where weak CO emission has been detected; the lane has been
suggested to be the funnel of accretion of unprocessed gas that is
feeding the central starburst \citep{turn97,meie02}.  Gas ionization
is centrally concentrated in this galaxy \citep{mart95,calz97,calz99},
and dominated in the optical by a 1--2~Myr old, and possibly younger,
super-star-cluster candidate heavily extincted by dust
\citep{calz97,crow99,trem01}. This super-star-cluster candidate may be
coincident with the brightest infrared source in the galaxy, as the
two are only slightly displaced from each other, about 5~pc in
projection \citep{turn03}. \citet {gorj01} place the age of the
super-star-cluster candidate at no more than a few hundreds of
thousand years, based on its infrared properties. The UV--detected
stellar clusters are systematically younger than $\sim$20~Myr
\citep{trem01,harr03}, and are distributed amid a typically older,
UV--bright, diffuse stellar population spread across the central
$\approx$350~pc \citep{meur95,calz97}. The characteristics of the
cluster and diffuse populations suggest that the clusters may be
dissolving into the diffuse population on timescales of a few tens of
Myr, due to tidal disruption, similar to what has been found in the
center of the Milky Way \citep{trem01}.  The velocity structure of the
ionized gas becomes disturbed beyond $\sim$200~pc radius from the
center, with velocity gradients of 10--30~km~s$^{-1}$, possibly a
tracer of slowly expanding gas \citep{mart95}. Ground--based images
show that gas ionized by mechanisms other than photoionization is
present at least between 560~pc and $\sim$1.4~kpc distance from the
galaxy center \citep{calz99}. This is in agreement with the presence
of two kpc--size superbubbles in the periphery of the ionized gas
emission, roughly along the major and minor axis, respectively, one of
which expanding with v$\sim$35~km~s$^{-1}$ \citep{marl95}.

\subsection{NGC~5236 (M~83)}

The central region of the massive, metal--rich spiral NGC5236 is the
site of a powerful starburst, although star formation is active across
the entire body of the galaxy, along the spiral arms. The central
starburst occupies a region about 400~pc across, and is bright at all
wavelengths, from the X-ray to the radio
\citep[e.g.,][]{ehle98,kinn93,roua96,tele93,turn94}. Two peaks of CO
emission, detected at the insertion points of the main stellar bar
onto the outer circumnuclear ring \citep{peti98,elme98,isra01},
suggest a picture in which the starburst is fueled by gas inflow along
the main stellar bar, collecting at the inner Lindblad resonance (see,
however, \citet{rega03}). In addition, a central minibar connecting
the outer circumnuclear ring to the inner circumnuclear ring may
provide a further channel for the gas to fuel the starburst
\citep{elme98}. The large amount of dust extinction, as inferred from
infrared data \citep{gall91,roua96} and from the large
infrared--to--blue ratio, is highly inhomogeneous, and multiple dust
lanes criss--cross the central region of the galaxy, alternating
heavily obscured with relatively unobscured regions. The population of
UV--bright clusters in the galaxy center has a mean age peaked around
5--7~Myr, and is distributed along the inner circumnuclear ring, in
the shape of a semi--circular ringlet between 50~pc and 130~pc from
the optical nucleus. Clusters younger than $\sim$5~Myr are located at
the edges of the ringlet, suggesting an outward propagation of star
formation, but also propagation from the southern area of the ringlet
towards the north \citep{harr01,puxl97}. Ground--based narrow--band
imaging of the center of the galaxy shows that the distribution of the
H$\alpha$ emission follows closely that of the blue stars and that
photoionization is the predominant ionization mechanism
\citep{calz99}; however, the coarse angular resolution of the
ground--based images (1$^{\prime\prime}$.3$\sim$30~pc) has prevented
establishing whether small--scale non--photoionized regions are
present within/around the starburst site.

\subsection{NGC~4214}

NGC4214 is a blue, low--metallicity, Wolf--Rayet Magellanic irregular
\citep{sarg91,kobu96}. The low extinction values measured from optical
spectroscopy \citep{maiz98}, together with the low infrared--to--blue
ratio, point consistently to a low dust content in the center of this
galaxy. Star formation is mainly active in two large complexes,
altogether spanning a size of $\sim$900~pc, more than twice the size
of the central site of star formation in NGC5253. A few additional
distinct actively star--forming regions are also present in the
central area of the galaxy, along the main bar
\citep{maiz98}. Multiwavelength observations indicate that star
formation may have started a few times 10$^7$~yr ago in this region
\citep{huch83}. NGC4214 shows a large variety of young ($<$10~Myr)
star forming complexes, from obscured, filled compact HII regions to
relatively unobscured, shell-like extended regions, which
\citet{mack00} interpret as an evolutionary/aging trend. The brightest
UV stellar knot in the starburst (NGC4214--1) is relatively young,
only $\sim$4--5~Myr old; it is located at the southern edge of an
H$\alpha$ cavity, and is likely to be density bounded
\citep{leit96}. Several shells of expanding gas surround the main
sites of star formation, with velocities between 30~km~s$^{-1}$ and
100~km~s$^{-1}$ \citep{mart98}.

\subsection{NGC~3077}

The gravitational interaction with M81 and M82
\citep{cott76,vand79,yunh94} is the likely trigger of the starburst in
the center of NGC3077. The HI morphology of this galaxy is 
heavily disrupted by the interaction, with an extended tidal arm to
the east of NGC3077 that contains about 90\% of the atomic gas of the
galaxy and is the site of a massive molecular complex
\citep{vand79,yunh94,walt99,walt02}. Similarly to NGC5253, star
formation is centrally concentrated, and its optical morphology is
heavily affected by patchy dust extinction. The dusty regions are
fairly well traced by a number of CO clouds, and their presence
produces an offset between the optical center and the infrared center
of the galaxy \citep{gull89,walt02}. The bright H$\alpha$ core is
surrounded by a number of expanding shells with velocities in the
range 40--100~km~s$^{-1}$ \citep{mart98}. The starburst hosts a number
of stellar clusters covering a fairly large range of ages, mostly
$\sim$1--100~Myr, with a cluster as old as $\sim$300~Myr
\citep{harr03}. This agrees, as order of magnitude, with the estimates
on the time since the last encounter of NGC3077 with M81, that
numerical simulations place at a few$\times$10$^8$~yr ago
\citep{brou91,thom93}.

\section{Observations and Data Reduction}

The images used in this work, all obtained with the Wide Field
Planetary Camera 2 (WFPC2) on-board the HST, are from a combination of
new observations (program IDs 8234 and 9144) and old or archival data
(program IDs 6124, 6524 and 6569), obtained during the period May 29,
1995 through July 18, 2001. A summary of the filters and exposure
times is given in Table~\ref{tbl-2}. The focus of this work is on the
ionized gas as probed by strong emission lines, thus the emphasis will
be on the narrow-band filter observations, although observations in
the medium/broad-band filters are also briefly discussed
(Table~\ref{tbl-2}), as they are used for stellar
continuum-subtraction or for comparing the blue stellar population (UV
filters) with the gas emission.

A subset of the emission-line data used here have been already presented
and discussed in previous papers. \citet{calz97} use the HST H$\alpha$
and H$\beta$ images of NGC~5253 to derive dust reddening
maps. \citet{mack00} discuss the morphology of the H$\alpha$ and
[OIII]~$\lambda$5007~\AA~ emission in NGC~4214. \citet{harr01,harr03}
present the HST H$\alpha$ images of NGC~5236, NGC~5253 and NGC~3077
for the purpose of constraining the ages of the stellar clusters in those
galaxies, and use the reddening maps derived in the present work. The
distinguishing characteristic of the present work relative to those
previous ones is the study of the interstellar gas ionization
mechanisms, rather than the investigation of the stellar population
content of the galaxies; thus, previous results will be used only as
necessary.  We add to previously published datasets images in the
narrow-band filters corresponding to the line emission in
[SII]~$\lambda\lambda$6726,6731~\AA~ (all galaxies),
[OIII]~$\lambda$5007~\AA~ (NGC3077, NGC5236, and NGC5253), and
H$\beta$~$\lambda$4861~\AA~ (NGC3077 and NGC4214). Also, although
\citet{harr01,harr03} presented H$\alpha$ images for NGC3077, NGC5236,
and NGC5253, these were not discussed in detail. Table~\ref{tbl-2}
gives the full list of narrow- and medium/broad-band imaging for
completeness. We performed anew the data reduction and combination of
those datasets that had not been previously presented by us, to ensure
homogeneity across the full suite of images.

Different programs used slightly different observing strategies, as
detailed in \citet{calz97,mack00,harr01,harr03}. We briefly recall
here those characteristics of relevance to the following analysis. The
centers of the four galaxies were imaged in the WF3 chip
(80$^{\prime\prime}\times$80$^{\prime\prime}$, NGC3077 and NGC5253),
in the PC1 chip (36$^{\prime\prime}\times$36$^{\prime\prime}$,
NGC5236), or straddling mainly the WF2 and WF3 chips (NGC4214). In all
cases, the selected WFPC2 chip contains the full central starburst
region. In particular, the WF3 or full--WFPC2 FOVs used for the three
dwarf galaxies, NGC3077, NGC4214, and NGC5253, subtend between 1.4~kpc
and 2~kpc of the central regions, with NGC4214 being probed on the
largest spatial scale (column~6 of Table~\ref{tbl-1}). In NGC5236, the
PC chip probes less than 1~kpc across the center (column~6 of
Table~\ref{tbl-1}), still sufficient to cover the extent of the entire
nuclear starburst and surrounding regions.

Program 9144, whose scope was to observe NGC3077 and to complete the
narrow-band datasets of NGC~4214 and NGC~5253, was implemented to
reproduce as closely as possible the pointings of the pre-existing
images. Typical rotations between narrow-band datasets for these
galaxies were small, $<$5$^o$. Images from programs 6124, 6524 and
9144 were dithered by a few pixels among exposures in the same filter,
to easily remove hot pixels during the data combination
steps. Programs 6569 and 8234 did not employ a similar observing
technique, and hot pixels were simply corrected for or flagged using
the IRAF/STSDAS task WARMPIX. Flagged data were not used in the
analysis. Observations of the emission lines used the same set of four
narrow-band filters for all four galaxies, as redshifts are small
enough that the lines fall within the filter bandpasses
(Table~\ref{tbl-1}). For the V--band continuum, programs 6524, 8234,
and 9144 used the medium band filter F547M, as its bandpass excludes
emission from strong lines, in particular the
[OIII]~$\lambda$5007~\AA; program 6569 used the wider F555W filter as
V--band continuum, and this case will be treated separately below.

The data, both new and archival, were reduced by the STScI calibration
pipeline, via the On-The-Fly-Reprocessing (OTFR, see
\citet{bagg02}). The OTFR uses the best calibration reference files
available at the time of retrieval from the HST Archive; most datasets
were retrieved between mid--2001 and early 2002. The calibration pipeline
performs basic steps such as flagging of bad pixels, A/D conversion,
bias and dark current subtraction, flatfielding. Post-processing steps
performed by us included: identification/correction of hot pixels
using the task WARMPIX, registration of each galaxy's dataset to a
common position using rotations and linear shifts; cosmic ray
rejection, hot pixel rejection (for dithered images), and combination
using the IRAF/STSDAS task CRREJ.

Absolute photometric calibrations were applied to the images using the
calibration keyword PHOTFLAM included in the image headers. The
zeropoint offsets (drifts) discussed in \citet{bagg98} imply
small corrections, $\sim$3\%~ or less, to our absolute photometry. The
effect of contaminant build-up onto the WFPC2 window is small or
negligible at the optical wavelengths, but can be significant in the
UV. Observations in the most contamination--sensitive filter, F255W,
were obtained about 23 days after decontamination, implying an
efficiency loss of 11\% (Calzetti et al. 1997). For observations in
the other filters, most of them obtained within 7 days of
decontamination, corrections are smaller, typically 1--2\% or less, with
only the F336W (NGC4214) needing as much as a 3.8\% correction.

Charge Transfer Efficiency (CTE) corrections can in principle be
severe in the narrow--band filters, where background and source counts
tend to be low \citep{whit99,dolp00}. In our case, the problem is
partially mitigated by the fact that the `background' is represented
by the smooth galaxy's stellar population, as all our sample galaxies
fill the WF or PC apertures. In the specific case of NGC4214, the
pointing location was specifically chosen to minimize CTE problems. In
order to evaluate the impact of CTE on our emission line measurements,
we examine the case of the F487N and F502N images of NGC5236. These
images/galaxy combinations are chosen because of the low CCD
efficiency in the blue, of the high-angular resolution of the
observations (PC chip), and of the low observed H$\beta$ and
[OIII]~$\lambda$5007~\AA~ surface brightnesses. All these facts
combine up to make the galaxy background ($\sim$0.55~DN/pix in F487N
and $\sim$0.84~DN/pix in F502N) and source counts some of the lowest
in our sample. Thus, this case represents a reasonable upper limit to
the impact of CTE corrections in our images. Using the formulae of
\citet{whit99}, the loss in signal at the leading readout edge (the
West side of the galaxy) is about 15\% and 13\% in F487N and F502N,
respectively; the loss at the trailing edge is expected to be
negligible, as our sources are extended \citep{ries00}. The line
emission is more concentrated than the stellar continuum emission, and
the leading edge is almost exclusively continuum; the faint line
emission at this edge will be excluded by our sensitivity cuts in the
final images. A few pixels inside the leading edge, the `background'
is already as high as $\sim$3~DN/pix, which implies a CTE loss of
about 7\%--8\% and 5\%--6\% in F487N and F502N, respectively, for a
typical source of 40~DN. To avoid undersubtracting the stellar
continuum, we privilege the continuum levels at the trailing edge when
performing the subtraction in the narrow--band filters (see
below). Although CTE is a systematic error, we quote the combined
`loss' in the F487N and F502N filters, about 10\%, as part of our
random uncertainty when discussing the [OIII]/H$\beta$ ratios, because
of the difficulty of controlling continuum subtration at that level of
accuracy.

A $\sim$4$\times$4~pixel area in the F555W images of NGC4214
corresponding to knot I-As ($\alpha$(J2000)=12:15:39.441,
$\delta$(J2000)=36:19:34.99, \citet{mack00}) is saturated. The F555W
image is used for the subtraction of the stellar continuum from the
F502N and F487N line filters. Uncorrected saturation leads to
undersubtraction of the continuum and overestimate of the line
strength in the saturated region. In the following analysis of
emission lines in NGC4214, the small saturated area will be neglected.
In NGC5253, the pixel corresponding to the peak of the F656N emission
($\alpha$(J2000)=13:39:56.016, $\delta$(J2000)=$-$31:38:25.05,
\citet{calz97}) is saturated even in the shortest exposure. Although
\citet{calz97} attempts at correcting for the saturation in
this pixel, we will ignore it in the following analysis.

Finally, all images were binned by 3x3 pixels
(0$^{\prime\prime}$.3$^2$ for NGC3077, NGC4214, and NGC5253, and
0$^{\prime\prime}$.14$^2$ for NGC5236), to smooth out small
registration shifts between images. The final bin size corresponds to
a physical scale between 3~pc and 5.8~pc, depending on the galaxy and
the chip used (column~7 of Table~\ref{tbl-1}). This scale is
comparable or larger than the typical half--light radii of young
stellar clusters in starbursts \citep{meur95,calz97}, but smaller than
the typical diameters of HII regions, these being in the range
$\approx$50--300~pc (\citet{oey03} and references therein); thus, the
binning still preserves high enough spatial detail to investigate
small--scale ionization characteristics.

\subsection{Derivation of the Emission Line Images}

\subsubsection{Subtraction of the Stellar Continuum}

The most delicate post-processing step in the present work is the
subtraction of the stellar continuum from the narrow--band images to
produce emission--line--only images.  This step is particularly tricky
for the H$\beta$ and [OIII] images of NGC3077 and NGC5236, owing to
the unfavorable combination of weak lines (due to the high
metallicities and large dust reddening in the galaxies,
Table~\ref{tbl-1}), and low detector quantum efficiency of the WFPC2
at blue wavelengths \citep{bire02}. For these two galaxies, as well as
for NGC5253, we verified the correcteness of the continuum subtraction
by comparing the WFPC2 line fluxes with the fluxes measured from the
large--aperture spectra of \citet{mcqu95} and \citet{stor95}.

The nebular emission lines contributing to the flux in each of the
four narrow--band filters are listed in column~2 of
Table~\ref{tbl-2}. Basically, each narrow--band filter is targeting a
single major emission line, with the exception of F656N, which
includes both H$\alpha$ and [NII]. The subtraction of the [NII]
contribution from the F656N flux will be discussed at length in 
section~3.1.2. No additional nebular lines are
expected to provide major contributions to the total flux in the
filters.

For NGC5253, stellar continuum images at the central wavelengths of
the narrow--band filters were obtained by linear
interpolation/extrapolation from the continuum F547M and F814W images
\citep{calz97}. Re-scaling factors of 10\% or less were then applied
to the continuum images to match the intensity of stars in the
narrow--band filters. The linear extrapolation of continuum images
could not be applied to NGC3077 and NGC5236, because of the
significant small-scale color variations of the stellar
populations. For these two galaxies, re-scaled F547M and F814W frames
were directly applied as continuum images to the F487N and F502N
images and to the F656N and F673N images, respectively; re-scaling
factors were again selected to match the intensity of stars in the
narrow--band frames. In NGC4214, the F555W filter contains
contributions from the [OIII]~$\lambda$5007~\AA~ and the H$\alpha$
emission; the first line is located on the blue side of the filter's
transmission curve, at 77\% of the peak, while the second line is in
the red wing, at 13\% of transmission peak. A recursive technique,
conceptually similar to that described in \citet{mack00}, is used to
recover a pure stellar continuum image by iterative subtraction of the
F502N and F656N images from the F555W image. The pure--stellar F555W
image and the F814W image were then linearly interpolated/extrapolated
to create the stellar continuum images at the appropriate wavelengths
for the narrow band filters.

The continuum--subtracted fluxes in the narrow--band images were
compared with fluxes from ground--based spectra for NGC3077, NGC5236,
and NGC5253 \citep{mcqu95,stor95}. The WFPC2 images were rotated to
the standard N--E position, and the flux contained in an aperture
corresponding to the spectral aperture was measured. The ground--based
spectra were convolved with the WFPC2 narrow--band filter's spectral
response, and continuum--subtracted. Line fluxes agree within
typically 10\% between images and spectra for H$\alpha$, and H$\beta$,
and within $\sim$15\% for [SII]. Discrepancies between images and
spectral measurements are generally in the same direction for all
lines in each galaxy, in the sense that for each galaxy the line
fluxes from images will be either systematically higher or
systematically lower than the fluxes from spectra. This characteristic
greatly mitigates discrepancies, for the purpose of analyzing line
ratios (e.g., [OIII]/H$\beta$ or [SII]/H$\alpha$). For the [OIII]
line, agreement between imaging and spectroscopy fluxes is more
variable, possibly owing to the large range of metallicities (and
thus, of [OIII] strength) covered by our galaxies. In the
low--metallicity, strong [OIII] emission galaxy NGC5253 the agreement between
images and spectra is about 12\%, which we consider good enough for
our purposes. In the high--metallicity, weak [OIII] emission galaxies NGC3077
and NGC5236, the [OIII] line fluxes in the images are about 30\% and
almost a factor of 2 higher, respectively, than the same lines in the
spectra. Because of their intrinsic weakeness, measurements of the
spectroscopic [OIII] lines in these two galaxies are highly uncertain,
and this may account for part of the discrepancy. Attempts to bring
the [OIII] fluxes in the NGC3077 and NGC5236 images into better
agreement with the spectroscopic fluxes, e.g., by increasing the amount of
stellar continuum subtracted from the narrow--band image, have not
been successful. The product is an obviously oversubtracted line
image. The line images adopted in the analysis are those corresponding
to the subtraction of the maximum continumm that does not produce
obvious oversubtraction.

The continuum--subtracted narrow--band images were then corrected for the
throughput values of the filter transmission curves at the observed wavelength
of the emission lines (Table~\ref{tbl-3}). Only in the case of the [OIII] and
H$\alpha$ emission in NGC5236 are these corrections as high as 10\% and
22.5\%, respectively, substantially more than the few percent values that
characterize all the other cases. NGC5236 is the galaxy with the highest
redshift in our sample, and the [OIII] and H$\alpha$ lines fall close to the
red wing of the filters. Such a large correction induces also a larger
uncertainty in the final fluxes, that will be folded in the subsequent
analysis. 

To further test the correcteness of our continuum subtraction in the
case of NGC5236, we checked the [OIII]/H$\beta$ and [SII]/H$\alpha$
ratios measured from our images against the small--aperture
spectroscopic results of \citet{dufo80} (for H$\alpha$, we first
subtracted the [NII] contribution, see next section). These authors
targeted the nucleus and a number of HII regions in the galaxy; the
line ratios we measure for the photoionized regions in the center of
NGC5236 are consistent with those authors' nuclear values.

In the case of NGC4214, we compared our H$\alpha+$[NII] and [OIII]
line photometry with the published values of \citet{mack00} for their
Regions~I and II. To prevent biases in the results from differences in
the centering of the apertures, we performed the comparison using only
the large apertures listed in Table~2 of \citet{mack00}, all of them
with radius $\sim$10$^{\prime\prime}$ or larger. For both lines, our
values are 26\%--28\% systematically lower than \citet{mack00}'s values,
independently of the Region used.  Since the data come from the same
original images (Table~\ref{tbl-2}), a possible explanation for the 
discrepancy is that we have performed a more stringent subtraction of 
the underlying continuum than those authors.  

To verify that such discrepancy does not impact our conclusions on
NGC4214, the [OIII]/H$\beta$ and [SII]/H$\alpha$ line ratios from our
images were compared with the spectroscopic ratios of \citet{mart97}
and \citet{maiz98}. Our line ratios occupy, indeed, similar loci in
the [OIII]/H$\beta$--versus--[SII]/H$\alpha$ and
[SII]/H$\alpha$--versus--H$\alpha$~intensity planes as those from
\citet{mart97} and \citet{maiz98}, after accounting for differences in
spatial resolution and depth between the images and the spectra. Thus,
our line emission images of NGC4214 can be considered adequate for the
purpose of the following analysis.

\subsubsection{Subtraction of the [NII] Emission Line}

The contribution of the [NII]~$\lambda$6548~\AA~ (plus
[NII]~$\lambda$6584~\AA~ in NGC3077) to the F656N images has been
removed using two different assumptions: (1) that the values measured
from the large-aperture spectroscopy are a good representation of the
average ([NII]$\propto$H$\alpha$); (2) that the [SII] line
emission map is a good tracer of the [NII] ([NII]$\propto$[SII]). The
second assumption stems from one main reason: the ratio [NII]/[SII]
has been observed to remain relatively constant, with variations of a
factor of 2 or less, in the diffuse ionized gas of a number of
galaxy haloes \citep{rand98,haff99,otte99,otte02}, far less than the
variation of [NII]/H$\alpha$. The fact that [NII]/[SII] is relatively
constant compared to [NII]/H$\alpha$ is due to two factors: (a) the
lower dependence of [NII]/[SII] on abundance and (b) the lower
sensitivity of [NII]/[SII] to the ionization properties of the extreme
UV radiation field \citep{kewl02}.  Thus, the assumption of a constant
[NII]/H$\alpha$ ratio is an over-simplification, which may impact the
H$\alpha$ flux determinations in low--excitation regions, where the
[NII]/H$\alpha$ ratios are expected to be high. We discuss here both
methods as a way to bracket extremes in the line ratio values, and
evaluate their impact on our conclusions.

The first assumption for the [NII]-subtraction is of straightforward
application to our data. For NGC4214, for which we do not have a
large-aperture optical spectrum, we have adopted the same
[NII]/H$\alpha$ ratio as in NGC5253; this is justified by the similar
metallicity and ionization conditions of the two galaxies
(Table~\ref{tbl-1} and section~4). Table~\ref{tbl-3} reports the
fractional contribution of [NII] to the F656N flux for each galaxy, as
calculated from the large aperture spectra. The only case in which the
[NII] correction to the F656N images is significantly above a few
percent is NGC5236, owing to its high metallicity. NGC3077 is also an
above--solar metallicity galaxy, with [NII]/H$\alpha\sim$0.33;
however, because of the low redshift of the galaxy, the two [NII]
lines fall in the wings of the F656N filter, providing a very small
contribution to the total flux.

For the second [NII]-subtraction assumption, we have generated `[NII]'
images that are proportional to the [SII] images, with the
proportionality constant derived from the ground--based spectra of
\citet{mcqu95} and \citet{stor95}. New H$\alpha$ emission line maps
have then been created after subtraction from the F656N images of the
`[NII] images'. The difference between these H$\alpha$ images and
those created under the first assumption is rather small, $<$2\%, for
NGC4214 and NGC5253, and $\le$3.5\% for NGC3077.  It is, however,
significant in NGC5236, where differences up to a factor 2.8 in the
H$\alpha$ fluxes of the low--ionization regions are observed between
the two methods. In the following, we adopt the H$\alpha$ maps
obtained from the [NII]$\propto$[SII] assumption as our default, but
recall the H$\alpha$ from the [NII]$\propto$H$\alpha$ assumption
whenever relevant to bracket the parameters' range in NGC5236.

Once H$\alpha$--`only' emission line images have been created, the
H$\alpha$ and H$\beta$ fluxes are corrected for the effects of the
underlying stellar absorption, in order to fully recover the nebular
fluxes. The weak H$\beta$ is more affected than H$\alpha$ by this
problem, but the images of both lines have been corrected. We have
adopted a value of EW$_{abs, H\alpha} \sim$EW$_{abs, H\beta}$=2~\AA~
for our galaxies \citep{mcca85}, which is representative of the values
measured in HII regions and starburst galaxies
\citep{mcca85,stor94}. The correction is performed as follows. The
emission line images are divided by the appropriate stellar continuum
image to create images of the line equivalent width (EW), for all bins
above a 5~$\sigma$ threshold. The bin values in each EW image are then
increased by 2~\AA~, and new line fluxes are computed based on the
revised EWs. The resulting emission line H$\alpha$ and H$\beta$ images
are used throughout this paper.

\section{Line Ratios}

Maps of the line ratios [OIII]/H$\beta$, [SII]/H$\alpha$, and
H$\alpha$/H$\beta$ are created for each galaxy from the line images,
after selecting only the data above the 5~$\sigma$ threshold in each
image (Table~\ref{tbl-3} for a list of the 1~$\sigma$ flux levels in
each line image). The threshold is chosen so to avoid that our results
are dominated by sources of random or systematic error.

The H$\alpha$/H$\beta$ maps are used to apply pixel-by-pixel dust
reddening corrections to fluxes, surface brightnesses, luminosities,
and the metal line ratios.  These corrections are typically small for
the metal line ratios, owing to the proximity in wavelength of
H$\beta$ to [OIII] and of H$\alpha$ to [SII]. For the same reason, our
simplistic assumption that the dust is located in a foreground screen
is not affecting these line ratios in a measurable way. However,
fluxes, surface brightnesses, and luminosities will be more influenced
by variations in the dust amount along the line of sight. In
particular, global values will be somewhat underestimated, because in
regions of high dust content the assumption of foreground dust will
provide insufficient corrections, and/or the hydrogen emission lines
will be undetected altogether. In what follows, the impact of dust on
these quantities and our results will be evaluated on a case-by-case
basis.

Line intensity and ratio maps for the four galaxies are shown in
Figures~\ref{fig1}--\ref{fig4}, panels b--d. Figure~\ref{fig5} shows
the histograms of the area occupied by different values of the metal
line ratios, [OIII]/H$\beta$ and [SII]/H$\alpha$, for each of the four
galaxies. The peak in each histogram marks the completeness limit of
the line ratio, as verified by the following
experiment. Histograms of the [SII]/H$\alpha$ line ratios were
produced at two different sensitivity thresholds, 5~$\sigma$ and
3~$\sigma$; as expected in the case of a selection effect
(completeness limit), the histogram peaks moved to higher values of
[SII]/H$\alpha$ for the deeper images. As an additional experiment,
the sensitivity cut was pushed to lower values by selecting a
3~$\sigma$ threshold in images binned by 5$\times$5~pixels (rather
than our adopted 3$\times$3~pixels bins); the histogram from these
deeper images reinforced the trend of the previous ones, thus
confirming the selection effect nature of the peaks. Despite this
effect, comparisons among the line ratio histograms of the four
galaxies are still meaningful, as similar depths are achieved between
images of the same emission line (Table~\ref{tbl-3}).

In the case of [SII]/H$\alpha$ (Figure~\ref{fig5}, left panel), the
value of the peak is similar for all four galaxies; this is due to the
low sensitivity of the line ratio to metallicity (Figure~\ref{fig6})
and to the hardness of the ionizing radiation field (for high
ionization parameter values, \citet{mart97}). Conversely, in the case
of [OIII]/H$\beta$ (Figure~\ref{fig5}, right panel) the position of
the peak changes from galaxy to galaxy, due to the high sensitivity of
this line ratio to the galaxy's metal content and radiation field's
hardness. For example, for fixed ionizing field and ionization
parameter, [SII]/H$\alpha$ varies by a factor $\sim$6 across the
entire metallicity range (Figure~\ref{fig6}); in the same conditions,
[OIII]/H$\beta$ varies by a factor $\approx$150 \citep{kewl01}. For
fixed metallicity, e.g. 20\% the solar oxygen abundance, a decrease of
25\% in the effective temperature of the ionizing source produces a
change of 30\% or less in the [SII]/H$\alpha$ ratio, and a far larger
decrease, a factor $\sim$2.5 or larger, in the [OIII]/H$\beta$ ratio,
for ranges of the ionization parameter typical of star--forming
regions ($-4.6 \le \log U \le -1.9$, \citet{mart97}). The latter
example helps explain the case of NGC4214 and NGC5253: the two
galaxies have almost identical metallicities but slightly different
peaks in the [OIII]/H$\beta$ histogram. Likely, the ionizing
sources in NGC5253 are harder, i.e., stars have higher effective
temperatures on average, than those in NGC4214. This could be
accomplished if photoionization were dominated by somewhat younger (by
a few Myr) stars or star clusters in NGC5253 than in NGC4214.

\section{Photoionized and Non--Photoionized Regions}

\subsection{Separating Photoionization from Other Mechanisms}

The \citet{bald81} and \citet{veil87} diagnostic diagrams provide an
excellent means of classifying galaxies according to their excitation
mechanism.  These diagrams are composed of easily measured line
ratios: log([OIII]/H$\beta$) line ratio on the y-axis against
log([NII]/H$\alpha$), log([SII]/H$\alpha$), or log([OI]/H$\alpha$) on
the x-axis.  Starbursts fall onto the lower left-hand region of these
plots, while narrow-line Seyferts are located in the upper right and
LINERs lie in the lower right hand zone.  The log([SII]/H$\alpha$)
ratio has traditionally been used as a diagnostic for shock-excited
gas and for the identification of supernova remnants
\citep[e.g.,][]{math72,phil98,dopi97}. Shock models predict that
relatively cool high-density regions form behind the shock front and
emit strongly in [SII] \citep{dopi78}.

Recently, \citet{kewl01} combined stellar population synthesis
modeling with the MAPPINGs~III photoionization and shock code
\citep{suth93} to produce starburst grids on the standard optical
diagnostic diagrams.  These grids were produced for a wide range of
metallicities (Z=$0.05 - 3.0\times$solar) and ionization parameters
($q=5\times10^{6} - 3\times10^{8}$cm/s, where $q$ is linked to the
nondimensional ionization parameter U by: U=$q/c$).  Using
these grids, \citet{kewl01} established an upper boundary on the
diagrams called the `maximum starburst line'.  The flux ratios of any
object lying above this boundary cannot be modelled by pure starburst
models alone.  Such ratios require the additional contribution by a
harder ionizing source like an AGN or shock excitation.  In a sense,
the use of this maximum starburst line is conservative because some
objects lying to the left and below the line may have a non-negligible
component of non-photoionization.  Indeed, the mixing line in
\citet{kewl01} shows that objects lying just to the left of the
maximum starburst line in the log([OIII]/H$\beta$) versus
log([SII]/H$\alpha$) diagram may have a contribution of up to
$\sim$30\% of non-photoionization to their emission. On the other
hand, the maximum starburst line is useful for our purposes because
any point in our sample lying to the right of the maximum starburst
line is very likely to be dominated by non--photoionization processes.

Comparisons with shock models confirm the above statement. The locus
of shock models by \citet{shul79} is shown in Figure~\ref{fig7},
denoted by an 'S'; these models include ionization from both the shock
and the photoionized precursor. The photoionized precursor is a region of
gas in front of the shock which has been ionized by the radiation
field emitted by the high-density gas at the shock front. A similar
locus is occupied by the shock models of \citet{alle99}. This lends
support to the fact that points lying above and to the right of the
maximum starburst line are likely to be dominated by shock emission or
other non--photoionization process. Although the maximum starburst
line was derived for discriminating the main ionization mechanisms in
`whole' galaxies, we apply it to our spatially--`resolved' analysis of
the excitation mechanisms within starbursts as a conservative approach
to the problem.

Using our data, we have therefore constructed diagnostic diagrams
log([OIII]/H$\beta$) versus log([SII]/H$\alpha$) for each galaxy
(Figure~\ref{fig7}, where the data points are bins from the line ratio
images); photoionization models from \citet{kewl01} at the appropriate
metallicity for each galaxy are reported in each panel, together with
the maximum starburst line.  

Despite our conservative approach, both NGC5236 and NGC5253 show clear
evidence for the presence of a number of regions dominated by
non--photoionization processes (Figure~\ref{fig7}), that is, lying
above the maximum starburst line. The situation is less clear for
NGC4214 and NGC3077, and two considerations should be kept in mind:
\begin{enumerate}
\item The 5$\sigma$ flux cut affects the locus of the bins on
Figure~\ref{fig7}.  We show NGC4214 and NGC3077 in Figure~\ref{fig8}
with 3$\sigma$ cuts.  It is clear that there are more data points in
the non--photoionized region of the diagram than there were in 
Figure~\ref{fig7}. This is expected as non--photoionized gas tends to 
occupy low surface brightness regions. A
similar increase in the non--photoionized data points is observed in
NGC5253 and NGC5236 when 3$\sigma$ cuts are chosen in their images.
\item The parameter range covered by the data points in
Figure~\ref{fig7} is limited by the comparatively shallow [OIII] and
H$\beta$ images; in the specific case of NGC3077 and NGC4214, the
images are probably not deep enough to extensively probe
non--photoionized regions, when using a 5~$\sigma$ sensitivity
cut. Indeed, the deeper [SII] and H$\alpha$ images show that a larger
fraction of bins have [SII]/H$\alpha \ge$0.7--0.9 for all four
galaxies (e.g., Figure~\ref{fig5}, left panel, and Figure~\ref{fig10})
than is apparent from the diagnostic diagram of Figure~\ref{fig7};
such high ratios are tell--tale signs of non--radiative processes
\citep{shul79,veil87,alle99}.
\end{enumerate}
For the purpose of our analysis, regions of non--photoionization
(Figures~\ref{fig1}--\ref{fig4}, panels~e) are defined to satisfy at
least one of the following two conditions: (1) the line ratios in the
regions are above/right of the `maximum starburst line'
(Figure~\ref{fig7}); (2) the [SII]/H$\alpha$ ratio is high enough to
be compatible with non--photoionization processes (Figure~\ref{fig10}).

For NGC5236, we investigate also the impact of the adopted
[NII]--subtraction method for the H$\alpha$ images on the
discrimination between photoionized and non--photoionized
regions. Figure~\ref{fig9} shows the effect on the diagnostic diagram
of each of the two assumptions: [NII]$\propto$[SII] (the default in
all Figures) or [NII]$\propto$H$\alpha$.  The first assumption clearly
causes more data bins to be located in the non--photoionized area of
the diagnostic diagram. Its main effect on the non--photoionzed
regions is to make them more extended, rather than `creating' new
ones. In addition, the two assumptions produce nearly identical
integrated H$\alpha$ fluxes, with the flux from the
[NII]$\propto$[SII] assumption being only 2\% smaller than with the
other assumption; this implies a negligible impact on the integrated
luminosities and derived quantities in Table~\ref{tbl-4}. The two
methods give similar results in terms of the {\em number and overall
morphology} of the non--photoionized regions, but other quantities
related to those regions, like the H$\alpha$ flux, the size, and the
areal fraction, do change and, in particular, they decrease in value
for the case [NII]$\propto$H$\alpha$. Although our default assumption,
[NII]$\propto$[SII], is the most physically justified (see discussion
and references in section~3.1.2), we report in Table~\ref{tbl-5} also
the numbers from the assumption [NII]$\propto$H$\alpha$, to highlight
the differences between the two methods for NGC5236 (see also
discussion below).

Non-radiative ionization is typically confined to regions of low
H$\alpha$ surface brightness, as already observed by other authors for
these and other galaxies
\citep{ferg96b,mart97,wang98,calz99,haff99}. Figure~\ref{fig10} shows
the distribution of values of [SII]/H$\alpha$ as a function of the
normalized H$\alpha$ surface brightness,
$\Sigma_{H\alpha}$/$<\Sigma>$. $<\Sigma>$ is defined as the mean
surface brightness at the half--light radius (\citet{wang98} and
Table~\ref{tbl-4}). The [SII]/H$\alpha$--versus--H$\alpha$ plot show a
relatively uniform behaviour across the four galaxies. The
photoionized gas, identified by the low--valued [SII]/H$\alpha$, spans
the full range of surface brightness in each galaxy (modulo the
sensitivity limits of our images), while the non--photoionized gas
crowds in the low~surface~brightness/high~[SII]/H$\alpha$ locus of the
diagram.  Conversely, the plot of [OIII]/H$\beta$ as a function of the
normalized H$\alpha$ surface brightness differs markedly from galaxy
to galaxy (Figure~\ref{fig11}). In NGC5236, the high
H$\alpha$~surface~brightness regions (HII regions) are characterized
by low values of the ratio [OIII]/H$\beta$, lower than the values
found in the low surface brightness areas. The trend is clearly
opposite in NGC4214 and NGC5253, where high surface brightness regions
have consistently high values of [OIII]/H$\beta$.  The datapoints of
NGC3077 show a trend that is in--between those of NGC5236 and
NGC4214/NGC5253. At the high end of the H$\alpha$ surface brightness
($\Sigma_{H\alpha}$/$<\Sigma>\gtrsim$3), there is a factor 60 or more
difference in [OIII]/H$\beta$ values between the four galaxies,
showing that the [OIII]/H$\beta$--versus--H$\alpha$~surface~brightness
plot is heavily affected by the galaxy's metallicity, with a clear
trend going from the metal--rich NGC5236, through NGC3077, down to the
metal--poor NGC4214 and NGC5253. At the same time, the plot shows more
uniform values at the low~surface~brightness end
($\Sigma_{H\alpha}$/$<\Sigma>\lesssim$0.5), where non--photoionized
gas is present; here, the [OIII]/H$\beta$ ratio spans a factor less
than 5 between the four galaxies.

Under our assumptions, in all four galaxies the fraction of H$\alpha$
emission contributed by mechanisms other than photoionization is
modest, around 3\%--4\% (Table~\ref{tbl-5}). In NGC5236, this fraction
becomes noticeably lower, about 1\%, when the [NII]$\propto$H$\alpha$
assumption, rather than the default [NII]$\propto$[SII], is
adopted. In order to extract the largest possible fraction of
non--photoionized gas from the images, we use the 3~$\sigma$ line
ratio images to derive the values reported in Table~\ref{tbl-5},
columns 2--5. If the 5~$\sigma$ line ratio images are instead used,
the numbers in columns~2--5 of Table~\ref{tbl-5} for NGC3077, NGC4214,
and NGC5253 drop by fractions that range between 30\% and 65\%, with
negligible changes in the case of NGC5236. The use of the 3~$\sigma$
images, rather than the 5~$\sigma$ ones, is preferred in this case for
a more meaningful comparison with predictions from models (see
section~6).

Most of the data points for all four galaxies are in the `photoionzed
gas' part of the diagnostic diagram (Figure~\ref{fig7}). The
low--[SII]/H$\alpha$ points of NGC4214 and NGC5253 are compatible with
the tracks for gas at the measured metallicity (and harder radiation
spectrum for NGC5253). 

For NGC3077, the low--valued [SII]/H$\alpha$ points have lower
[OIII]/H$\beta$ than expected from models at the galaxy's metallicity
(left-top panel of Figure~\ref{fig7}); a track closer to the observed
data points is the one at Z=2~Z$_{\odot}$, with gas density
n=350~cm$^{-3}$. However, even this track does not fully account for
the distribution of the data points in NGC3077, especially those
located below the track itself. A starburst population comprising of
ionizing stars in a range of evolutionary stages, and a combination of
high-- and low--ionization components, may account for the dispersion
of the datapoints in the low--[OIII]/H$\beta$/low--[SII]/H$\alpha$
part of the diagram \citep{moy01}.  The Z=2~Z$_{\odot}$,
n=350~cm$^{-3}$ track is located well below the maximum starburst line
(Figure~\ref{fig7}); if the track itself is used as a boundary to
separate photoionized gas (below--left of the track) from
non--photoionized gas (top--right of the track) in NGC3077, we get
that the H$\alpha$ emission of the non--photoionized gas is 26\% of
the total. This fraction is 6 times higher than what we calculate
under the `standard' assumptions above, and it is also over 4 times
larger than what the starburst itself can support if all its
mechanical energy is used to shock the gas (Table~\ref{tbl-6}). Thus,
the distribution of the data points in the
[OIII]/H$\beta$--versus--[SII]/H$\alpha$ diagram for NGC3077 appears
to be due to a combination of multiple processes, which possibly
include the high-- and low--ionization components of \citet{moy01}, in
addition to a non--photoionized component.

In NGC5236 we find an even more extreme behavior than in the case of
NGC3077. While the low--[SII]/H$\alpha$ points are in the general
region of the high--metallicity theoretical tracks for photoionized
gas, they do not follow any specific track in this galaxy (top--right
panel in Figure~\ref{fig7}). The qualitative trend of the data does
not change if the H$\alpha$ images are corrected for [NII] using the
assumption [NII]$\propto$H$\alpha$, rather than [NII]$\propto$[SII]
(Figure~\ref{fig9}). One possible interpretation is that, in addition
to the presence of a complex age distribution in the starburst stellar
population and of regions of low-- and high--ionization \citep{moy01},
non--photoionized processes heavily affect the weak metal line ratios
in this metal--rich galaxy \citep[e.g.,][]{mart97}. Alternatively, the
emission may result from a shock--only component (without a
photoionized precursor), as the ones modelled by \citet{alle99}. This
can occur if the shock is propagating into a low density environment
like the neutral interstellar medium, or if we are in the presence of
low--velocity shocks. Thus, the difference in line emission
characteristics between the giant spiral and the three dwarfs may
arise from the uniqueness of nuclear environments. If the
Z=2~Z$_{\odot}$, n=350~cm$^{-3}$ track is again used as a boundary
between photoionized gas and non--photoionized gas, the fraction of
H$\alpha$ emission associated with the non--photoionized component
increases to $\sim$9\%, about three times larger than what derived
with our `standard' assumptions. This more extreme scenario is still,
although barely, energetically supportable by the current starburst
(Table~\ref{tbl-6}). Morphologically, in the more extreme scenario the
two regions~A and B merge together and region~A gets extended along
areas of low H$\alpha$ surface brightness towards the nucleus of
NGC5236.

In summary, even under less restrictive assumptions for the
identification of non--photoionized gas, the photoionized gas still
accounts for the majority of the H$\alpha$ emission, at the level of
at least $\sim$75\% and $\sim$90\% in NGC3077 and NGC5236,
respectively.

Not surprisingly, the geometry of the photoionizing sources is more
complex than just a simple central point source, in all four
galaxies. This is easily seen in Figure~\ref{fig5}, left panel, where
the expected trend of areal coverage as a function of [SII]/H$\alpha$
is shown for a central point--like source; the line ratio in the case
of a central source is a function of the radial distance R from the
source itself, [SII]/H$\alpha\propto$R$^2$. In all cases, the observed
areal coverage exceeds the prediction from the single central source,
clearly showing that the ionizing sources are distributed across the
entire starburst region, as already evident from the UV images of the
same galaxies (Figures~\ref{fig1}--\ref{fig4}, panels~a,
\citet{harr01,harr03}). In particular, deviations from the single
point-like source geometry happen at radii smaller than 17--36~pc for
the four galaxies, where the radii increase from NGC5236, to NGC4214,
NGC5253, and NGC3077; this fact is compatible with the presence of
multiple ionizing clusters distributed within the starburst region
with intracluster distances typical of the sizes of HII
regions. Indeed, each `transition' radius is much smaller than the
radius of the Stromg\"en sphere and of the UV half--light radius
(Table~\ref{tbl-4}, even after the UV half--light radius is corrected
for residual effects of dust obscuration, see section~6.3), the latter
confirming the spatially extended nature of the starburst population
in each galaxy.

\subsection{Morphology of the Non--Photoionized Regions}

The morphologies of the non--photoionized regions can offer clues as
to the mechanism underlying the observed ionization. Here we review
the various structures observed in the four sample galaxies. They show
two basic morphologies: cavities or `enclosed' regions and shell (with
more or less filamentary structure) regions.

The cavities (e.g., regions A and B in NGC5236, Figure~\ref{fig3}, and
region~As in NGC4214, Figure~\ref{fig2}) are areas of low H$\alpha$
surface brightness surrounded by actively star--forming knots or
regions, the latter with marked high H$\alpha$ surface brightness. The
radii of the non--photoionized regions range, in our cases, from
$\sim$20~pc to $\sim$50 pc.

The H$\alpha$ shells surround the central starburst, and have more or
less marked filamentary structure. They often trace the closest edge
of more external bubbles or arcs to the central peak of
star--formation. In our three dwarf galaxies, these regions are
located at a radius of $\sim$0.2~kpc from the area of peak
star--formation. In NGC3077 and NGC5253, these radii are a factor
$\sim$2 larger than the H$\alpha$ half--light radii
(cf. Table~\ref{tbl-4} with Table~\ref{tbl-5}), and are also larger
than the UV half--light radii or the Stromg\"en radii, even after
accounting in the latter for the uncertainties in the filling factor
and gas density. In NGC4214, the size of the non--photoionized shell
is smaller than both the H$\alpha$ and the UV half--light radii.

In NGC5236 there is one additional concentration of
non--photoionized gas (region~C), located at the periphery of the
central starburst, on the southern side.  This region appears to
coincide with the point of insertion of the main stellar bar onto the
outer circumnuclear ring \citep{elme98}. Interestingly, there
is no similarly obvious non--photoionization region close to the other
insertion point (i.e., close to the other CO region as marked in
Figure~\ref{fig3}, panel~e). Likely, the extinction due to the dust
lane that dominates this other area has prevented a detection.

The PC chip image of NGC4214 contains a non--photoionization region
with a markedly annular geometry, of physical radius of $\sim$40~pc
and thickness of $\sim$20~pc.  This H$\alpha$ ring is located in a
fairly isolated area, devoid of other H$\alpha$ emitting regions and
is not surrounded by knots of active star formation, as the other
cavities discussed above. It is also located in a region of lower
stellar density than the central starburst. Within the ring, the
[OIII] emission is more centrally concentrated than the [SII]
emission, and the line ratio values are: [SII]/H$\alpha\sim$0.9--1
across the $\sim$90~pc covered by the ring; [OIII]/H$\beta\sim$1--3
within the inner $\sim$20~pc radius, with [OIII] undetected on larger
scale. The total H$\alpha$ luminosity associated with the ring is
4.3$\times$10$^{37}$~erg~s$^{-1}$, as measured from our images. We
believe this structure to be a candidate supernova remnant located in
the outskirts of the starburst in NGC4214, with coordinates
$\alpha$(J2000)=12:15:42.436 and $\delta$(J2000)=$+$36:19:47.34.  The
measured size, total H$\alpha$ luminosity, and line ratios are not
untypical of SNRs \citep[e.g.,][]{blai81,chuk88,blai94}. For an inner
radius of $\sim$20~pc and an adopted shock speed of 100~km~s$^{-1}$,
the Sedov-Taylor solution predicts an age of $\sim$8$\times$10$^4$~yr
for the candidate SNR \citep{oste89}.

\section{Discussion}

For three of the four galaxies in our sample, NGC3077, NGC4214, and
NGC5253, evidence for the presence of non--photoionized gas had
already been found by the spectroscopic and kinematic studies of
\citet{mart97,mart98}. However, because of the limited areal coverage
of the spectroscopic study, and of the absence of metal--line
information in the kinematic study, those previous works could not
quantify the prominence and extent of the non--photoionized gas within
the starburst.  These limitations are overcome by the present
investigation, that combines metal-- and hydrogen--line
high--resolution imaging data to identify and quantify
non--photoionized gas.

The comparable depths reached by the emission line images and the
relatively small difference in distance between the four galaxies
allow us to compare them in a statistically meanignful fashion.

Despite their different morphologies, there is a common trend in the
observed non--photoionized regions; they tend to show up as extended,
coherent structures, rather than, for instance, small individual
`blobs' as may be expected from multiple, uncorrelated supernova
remnants. This suggest that the underlying mechanism responsible for
the non--photoionized gas results from collective processes, such as 
multiple supernova explosions occuring over short timescales in spatially 
correlated or `grouped' stellar clusters.

In all four of our sample galaxies, the non--photoionization
mechanisms are responsible for only a small fraction of the projected
H$\alpha$ emission, around 3\%--4\% of the total detected in the WPFC2
images (Table~\ref{tbl-5}). This should be regarded to some extent as
a lower limit, because of the conservative approach we have adopted in
identifying non--photoionized gas. Regions marked as `photoionized'
may still contain a fraction of emission which originates from
non--radiative processes and which can substantially increase, e.g.,
by a factor of a few in our metal--rich galaxies, the H$\alpha$ flux
fraction associated with non--photoionized gas (section~5.1).  The
areal coverage of non--photoionized gas in the four galaxies is
between 15\% and 25\% of the total H$\alpha$ area, thus much larger
than the fraction by flux, as emission due to processes other than
photoionization is associated in general with regions of low H$\alpha$
surface brightness (Figure~\ref{fig10}).

Even making allowances for the fact that our observations may not
cover the full extent of the diffuse H$\alpha$ emission in the four
galaxies, we do not expect the H$\alpha$ flux fraction of
non--photoionized gas to increase significantly if wider--field
observations were employed. The non--photoionized gas is found in
regions of low H$\alpha$ surface brightness, and including more areal
coverage is not expected to increase the overall flux fraction by large
amounts (see also the discussion on the ground--based observations of
NGC5253 at the end of section~6.1). However, the areal fraction
occupied by the non--photoionized gas may increase significantly if
wider--field observations were employed.

Not all low H$\alpha$ surface brightness regions coincide with
non--photoionized areas; notable exceptions are areas of high absorption
due to large concentrations of dust. Here the H$\alpha$ surface
brightness is low because the strong dust absorption prevents an
accurate correction of the line and stellar continuum fluxes for its
effect. 

In this category are the main dust lane in the center of NGC5253 and
the gas and stellar `holes' in the center of NGC3077. Where the
signal-to-noise is sufficient, the line ratios in these areas are
consistent with photoionization of the gas. These
high--dust-absorption areas also coincide with the location of the
detected CO emission in both NGC3077 and NGC5253
(\citet{walt02,turn97,meie02}; see Figure~\ref{fig1} and
Figure~\ref{fig4}, panels~e). Thus, as expected, CO emission is
associated with young regions of star formation, and photoionization
of the ISM, while non--photoionization is displaced relative to these
regions.

A similar conclusion holds for the CO emission in NGC4214. The
H$\alpha$ cavity in this galaxy is `sandwiched' between the two main
CO peaks detected by \citet{beck95} (our Figure~\ref{fig2}, panel~e);
here, too, the CO emission corresponds to an area where
photoionization dominates, although, for one of the two peaks, it also
corresponds to relatively high H$\alpha$ surface brightness. A
parallel analysis cannot be performed for the nuclear region of
NGC5236, as the two peaks of CO emission \citep{sofu94,isra01} are
located outside the detected line ratio areas in our images.

\subsection{The Energy Balance of the Non--Photoionization Component: The Case for Shock--Ionization}

As mentioned in the Introduction, we are favoring ionization from
shocks and their precursors as a viable mechanism to produce the
non--photoionization component observed in the four galaxies
\citep{mart97}. This is different from the mechanisms invoked for the
haloes \citep{rand98,haff99,otte99,otte02} or the diffuse medium of
more quiescently star--forming galaxies \citep{hunt90,ferg96a}, but
there are some important differences between these galaxies and our
starbursts. The non--photoionized emission in our starburst galaxies
is more compact, covering the range $\sim$200--1,000~pc, than that in
galaxy haloes ($\sim$1--5~kpc). The typical H$\alpha$ surface
brightness of the non--photoionized regions discussed in this paper is
in the range
$\approx$5$\times$10$^{-16}$--2$\times$10$^{-14}$~erg~s$^{-1}$~cm$^{-2}$~arcsec$^{-2}$
(Figure~\ref{fig10} and Table~\ref{tbl-4}), or about 20--100 times
higher than the typical surface brightness of the diffuse ionized
medium in more quiescent galaxies \citep{wang98}, and higher by a much
larger factor than the DIG emission in galaxy haloes. The half--light
radius H$\alpha$ surface brightness $<\Sigma>$ is also 1--3~orders of
magnitude higher in the starbursts than in the more quiescently
star--forming galaxies \citep{wang98}.

Furthermore, X--ray observations are available from ROSAT, Chandra, or
XMM-Newton for all four galaxies
\citep{bi94,mart95,stri99,sori02,hart03}, and show presence of
extended soft X--ray emission in their centers over the scales of
interest (a few~tens of arcsecs, up to $\sim$1.4$^{\prime}$ in
NGC4214). In all cases, this extended X--ray component has been
attributed to thermal plasma emission from hot, diffuse gas. These
arguments provide circumstantial evidence that the mechanism for the
non--photoionized component in our starbursts is shock heating, rather
than the unkown ``extra heating'' process invoked for the very dilute
ionized gas in galaxy haloes or the diluted photoionization mechanism
used to explain the diffuse medium of more normal galaxies.

A basic check needs to be performed to verify whether the current
starbursts can provide the mechanical energy output necessary to
produce the observed non--photoionized H$\alpha$ emission.

The H$\alpha$ emission expected as a result of the starburst's
mechanical output, L$_{H\alpha,mech}$, is derived adopting the
\citet{bine85} prescription: L$_{H\alpha,mech}\sim$0.025~L$_{mech}$,
where L$_{mech}$ is the mechanical luminosity produced by the
starburst.  To derive the latter, we adopt the Starburst99 models
\citep{leit99}. In particular, each starburst is assumed to be well
described by continuous star formation in the range 10--100~Myr, and
Salpeter stellar Initial Mass Function up to 100~M$_{\odot}$. Each
galaxy is matched to the model closest in metallicity value. The
number of ionizing photons predicted by the Starburst99 models are
rescaled to match the observed photoionized H$\alpha$ luminosity
(Table~\ref{tbl-4}, after subtraction of the non--photoionized
component, see also column~2 of Table~\ref{tbl-6}). The rescaling
factors are then used to derive the expected mechanical luminosities
L$_{mech}$ for each starburst, and, finally, the expected
L$_{H\alpha,mech}$ and their fraction to the total H$\alpha$
luminosity (Table~\ref{tbl-6}).  The values of
L$_{H\alpha,mech}$/L$_{H\alpha}$ listed in Table~\ref{tbl-6} represent
therefore the expected fractions of H$\alpha$ luminosity that the four
starbursts can produce via shocks triggered by massive star winds and
supernovae, and should be compared with the observed non--photoionized
H$\alpha$ luminosity fractions measured from our 3~$\sigma$ images
(Table~\ref{tbl-5}, column~4). We prefer the use of the 3~$\sigma$
images for this part of the analysis, instead of our default
5~$\sigma$ images, as higher sensitivity cuts tend to exclude low
surface brightness regions, and, therefore, exclude preferentially the
non--photoionized areas in the images. A quantification of how much
non--photoionization is excluded by the 5~$\sigma$ images is given in
section~5.1. Therefore, the 3~$\sigma$ images can place more stringent
constraints on the ability of the current starbursts to sustain the
measured L$_{H\alpha,nph}$.

In all cases, the current starburst can support the observed
non--photoionized H$\alpha$ luminosities, implying that the mechanical
energy input into the ISM from supernovae and high--mass star winds is
sufficient to produce the level of non--photoionization observed in
all four galaxies, if star formation has been constant over the last
$\sim$few~10$^7$~yr. In the case of the massive galaxy NGC5236, a
starburst as young as 10~Myr can already account for the observed
L$_{H\alpha,nph}$, and an older starburst could produce even larger
mechanical H$\alpha$ luminosities. For the three dwarf galaxies, the
observed L$_{H\alpha,nph}$ fall closer to the high range of predicted
values L$_{H\alpha,mech}$ from models. In the Starburst99 models, the
ratio of the mechanical energy output to the number of ionizing
photons (and therefore to the photoionized H$\alpha$ emission),
L$_{mech}$/N$^o_{ion}$, reaches a constant value after
$\sim$3$\times$10$^7$~yr, for constant star formation. This implies
that the starbursts in the three dwarfs need to have produced stars at
the same rate as (or higher than) the present one for at least 30~Myr,
to account for the observed L$_{H\alpha,nph}$. This is likely to be
the case for our dwarf galaxies (section~6.4). Overall, shocks are a
viable mechanism, from an energetic point--of--view, for the observed
luminosity of the non--photoionized gas in the center of all four
galaxies.

The near--UV emission from the galaxies probes the radiative output
from stars more massive than $\sim$5~M$_{\odot}$, and thus can be used
as a tracer of the mechanical output from the starburst on timescales
of a few hundred million years (with assumptions on the
star--formation history), much longer than the `instantaneous' output
traced by the H$\alpha$ emission.  The reddening--corrected near-UV
fluxes, luminosities, and corresponding mechanical
luminosities of the four galaxies predicted by the Starburst99 models are
listed in Table~\ref{tbl-6}.

For all galaxies, L$_{H\alpha,mech}$ derived from the near--UV
emission is in very good agreement with the same quantity derived from
the photonionized H$\alpha$, despite uncertainties in the dust
extinction correction of both the UV and H$\alpha$ images (especially
in the high--opacity CO regions). To estimate the potential impact of
the latter, we consider the cases of NGC5253 and NGC3077. The basic
premise is that heavy dust extinction will preferentially affect the
central, photoionization--dominated regions of the starbursts. 

Dust opacity indicators, such as the UV slope and the infrared-to-UV
or infrared-to-blue ratios \citep{calz01}, give average UV
attenuations A$_{2600}\sim$1.1~mag and A$_{3000}\sim$1.4~mag, for the
starburst regions of NGC5253 and NGC3077, respectively; these are
factors 2 and 2.2 higher, respectively, than what we derive directly
from our H$\alpha$/H$\beta$ images. Using the same dust diagnostics,
the average attenuation at H$\alpha$ in the centers of NGC5253 and
NGC3077 is a factor 1.4 and 1.8 higher, respectively, than what
estimated from our optical images.  Even if all the `extra' dust
attenuation is associated with the photoionized gas for both galaxies,
the revised numbers for the intrinsic L$_{H\alpha}$ and, therefore,
for the expected L$_{mech}$, provide a L$_{H\alpha,mech}$ luminosity
that is at most 6\% and 13\% of the total {\em measured} H$\alpha$
luminosity in NGC5253 and NGC3077, respectively. This fraction will
decrease if any of the `extra' dust is associated with the
non--photoionized regions as well, and will converge to the
L$_{H\alpha, mech}$/L$_{H\alpha}$ values of Table~\ref{tbl-6}
(column~4) in the limiting case of an homogeneous distribution of the
`extra' dust among the photoionized and non--photoionized regions.

In summary, the mechanical energy output from the starbursts is
sufficient to support the observed L$_{H\alpha,nph}$ as measured in
section~5.1 (Table~\ref{tbl-5}). Furthermore, a considerable fraction
of this mechanical energy, between 70\% and 100\% (Table~\ref{tbl-6}),
needs to be used.

To support this conclusion, the large--scale H$\alpha$ emission (both
photoionized and non) needs to be included in the full budget
accounting. For this purpose, we combine the present results for
NGC5253 with those of \citet{calz99}. Our HST images and their
ground--based images probe complementary regions in this galaxy in
terms of non--photoionization (the HST images being at higher angular
resolution, but shallower than the ground--based images). This implies
that the non--photoionized shell in NGC5253 extends from 0.22~kpc all
the way to $\sim$1.5~kpc from the center of the galaxy. The sum of the
non--photoionized H$\alpha$ from this work and that of Calzetti et
al. gives L$_{H\alpha,nph}$=1.6$\times$10$^{39}$~erg~s$^{-1}$; from
that same work, the extinction--corrected H$\alpha$ integrated flux is
2.9$\times$10$^{-11}$~erg~s$^{-1}$~cm$^{-2}$, implying
L$_{H\alpha,mech}$=(0.012--0.038)~L$_{H\alpha}$=(0.68--2.10)$\times$10$^{39}$~erg~s$^{-1}$,
in good agreement with L$_{H\alpha,nph}$. Incidentally, even on this
extended spatial scale,
L$_{H\alpha,nph}$/L$_{H\alpha}\sim$0.03. Therefore, the measured
luminosity of the non--photoionized gas is still perfectly supportable
by the current starburst. We can expect that a similar argument holds
for NGC3077.

\subsection{The Nature of the Cavities in NGC5236 and NGC4214}

NGC5236, the brightest and most massive galaxy in the sample, is also
the only one that does not show evidence for extended H$\alpha$
structures associated with non--photoionized gas (Figure~\ref{fig3},
panel~e). The non--photoionized gas is mainly concentrated in two
localized regions surrounded by star-forming areas, except for a
sourthen concentration which corresponds to the insertion of the
stellar bar into the outer circumnuclear ring (section~5.2). More
generally, \citet{calz99} found, based on ground--based images, that
there is little or no evidence for diffuse ionized gas extended beyond
the regions occupied by the ionizing stars. Their conclusion is
supported by our data. The central starburst in NGC5236 is somewhat 
more compact than the starbursts in NGC5253 and NGC3077, as inferred
from both the H$\alpha$ and UV half--light radii (Table~\ref{tbl-4}),
and is only 15\% more powerful than the starburst in NGC5253 (see the
star formation rates in Table~\ref{tbl-4}). If comparable starbursts
produce extended gas structures of comparable sizes, we may expect
such structures to be located at a distance $\sim$200--250~pc from the
nucleus of NGC5236 (Table~\ref{tbl-5}). Our PC observations are
sensitive enough and sample enough region, $\sim$750~pc, to detect
extended gas structures in NGC5236, if present
(Table~\ref{tbl-1}). However, like \citet{calz99}, we do not detect
ionized gas extended beyond the region occupied by the starburst
stellar population. Those authors concluded that the absence of an
extended ionized gas component is evidence of the gas confinement
exerted by the deep potential well in the center of
NGC5236. \citet{calz99} did not detect any non--photoionized gas in
the center of the galaxy, probably due to the lower angular resolution
of the ground--based images. The fact that we do detect localized
non--photoionized gas in NGC5236, while still failing to detect a gas
component extended beyond the starburst stellar population, is further
support that the starburst in this galaxy is a local event and has
little influence on the large--scale galaxy's ISM.

The H$\alpha$ luminosity associated with the non--photoionized gas in
the two cavities in NGC5236 is 5.4$\times$10$^{38}$~erg~s$^{-1}$ for
region~A and 1.4$\times$10$^{38}$~erg~s$^{-1}$ for region~B,
accounting for 58\% of the total non--photoionized H$\alpha$ emission
in the center of the galaxy. These fairly large energy requirements
imply contributions from multiple supernova explosions. Indeed, they
correspond to the input mechanical luminosity of stellar clusters with
masses $\sim$5.5$\times$10$^5$~M$_{\odot}$ and
$\sim$1.5$\times$10$^5$~M$_{\odot}$ for region~A and region~B,
respectively, for a Salpeter IMF in the 1-- 100~M$_{\odot}$ range
\citep{leit99}. For expansion velocities between 50~km~s$^{-1}$ and
100~km~s$^{-1}$, the dynamical age of the cavities comes to
3--7$\times$10$^5$~yr and 1--2.5$\times$10$^5$~yr, respectively
\citep{mart98}. The dynamical ages of the cavities are thus small
relative to stellar evolution times. If those cavities are actually
due to mechanical input from a single cluster each, such clusters
would be a few times more massive than the current population of young
stellar clusters in the center of NGC5236 \citep{harr01}, but still
within the mass range of young stellar clusters in starburst galaxies
\citep[e.g., ][]{chan03}. Alternatively, a few clusters in each region
could be responsible for the powering of the cavities; the structure
of the starburst in this galaxy already provides morphological
evidence for young stellar clusters to amass in groups of 2--4 in
localized regions of projected size 30--40~pc \citep{harr01}. These
sizes fit confortably within the area occupied by each of the two
cavities.

The well--known central H$\alpha$ cavity in NGC4214
\citep{hunt82,leit96} is edged to the south by the brightest H$\alpha$
knot in the galaxy. However, the brightest UV cluster, NGC4214-1, does
not coincide with the bright H$\alpha$ knot, but is located to the NE
of it, within the cavity \citep{leit96}. The diameter of the cavity is
$\sim$7$^{\prime\prime}$ diameter, or $\sim$100~pc; this is about 50\%
larger than the size of the non--photoionization region detected in
our images (diameter$\sim$70~pc, Table~\ref{tbl-5}). The discrepancy
is likely to be physical: the edges of the cavity may be dominated by
the photoionization that is induced, possibly, by NGC4214-1. This
stellar cluster, at 4--5~Myr of age \citep{leit96}, is old enough to
have experienced the first supernova explosions (that start around
$\sim$3.5~Myr, \citet{leit99}), and thus to be the cause of the
detected non--photoionized gas within the cavity. The cavity's
dynamical age is $\sim$0.4~Myr, adopting the velocity v=70~km~s$^{-1}$
measured in \citet{maiz99}; the stellar cluster was at most 0.5~Myr
younger when the bubble started to form, old enough to have
experienced $\sim$200--700 supernova explosions. The amount of
non--photoionized H$\alpha$ associated with the cavity is
3.0$\times$10$^{37}$~erg~s$^{-1}$, implying a mechanical luminosity of
1.2$\times$10$^{39}$~erg~s$^{-1}$. If the stellar cluster was only
3.5--4.0~Myr old when the cavity started to expand, its required mass
to drive the estimated amount of mechanical luminosity is
3.8$\times$10$^4$~M$_{\odot}$, for a Salpeter IMF in the 1--
100~M$_{\odot}$ range. This is about a factor 4--5 lower than the
masses estimated by \citet{maiz99} and inferred from the 2200~\AA~
luminosity of \citet{leit96}, after rescaling for the different
adopted distances for NGC4214. However, our mass estimate is a {\em
minimum requirement} for producing the observed non--photoionized
H$\alpha$ luminosity, and cannot account for the fraction of the
mechanical energy lost to the rupturing the bubble \citep{maiz99}.

\subsection{The Shells in the Dwarf Galaxies}

Ionized gas taking the form of extended shells with more or less
filamentary structure is a common feature of dwarf galaxies
\citep{hunt92,hunt97}. Such shells are generally around 10~Myr old or
younger, with very few cases laying in the range 10--20~Myr
\citep{mart98}.

In the case of NGC3077 and NGC4214, the convex edges of the shells are
well traced by non--photoionization in our images
(Figures~\ref{fig1}--\ref{fig2}, panel~e). For NGC5253, our results
are less clear (Figure~\ref{fig4}, panel~e), but they provide a
picture similar to that of the other two galaxies, once combined with
the results of \citet{calz99}. Indeed, the inner size of the
non--photoionized shell we measure in NGC5253 coincides with the scale
at which \citet{mart95} find disturbed velocity structure in the
galaxy. In all cases, the shells could be outward starburst--driven
shocks (see discussion in section~6.1, and \citet{mart98}), if the
starbursts have been lasting for more than $\sim$30--50~Myr.
Interestingly, in all three cases the inner edge of the shell is
located at a radius of $\sim$0.2~kpc from the center of the
starburst. 

Another common feature among the three dwarfs is the fact that the
high-surface-brightness regions of the H$\alpha$ filaments are
generally photoionized, in agreement with previous findings
\citep{hunt94,hunt97}, while non--photoionized gas is preferentially
located at the inner or outer edges of the filaments. This is
especially evident in NGC3077 (Figure~\ref{fig1}, panel~e). The data
thus suggest that the optical emission from shocks, which is
relatively faint, dominates when the gas material is moving into low
density surroundings, but becomes a small fraction of the total
emission when the shock encounters regions of higher gas density, as
here the gas emitting from the photodissociation/photoionization edge
is likely to dominate. If this is true in general, the only shock
emission detected is the one located in low density regions, and is
therefore a lower limit to the total shock emission.

Both in NGC3077 and NGC5253, the inner size of the shells of
non--photoionization exceeds by $\sim$40\% the near--UV half-light
radius of the stellar population as measured directly from our
reddening--corrected data (Tables~\ref{tbl-4}--\ref{tbl-5}). Residual
dust obscuration in the UV images will have the effect of yielding
measured near--UV half--light radii that are larger that the actual
ones, if the dust is centrally concentrated. To quantify this effect,
we use the estimates on the UV dust attenuation given in
section~6.1. If, for both galaxies, we locate the `missing' UV flux
within the measured half--light radii, these would be reduced to
$\sim$84~pc for NGC3077 and to $\sim$110~pc for NGC5253. Thus, under
these assumptions, the UV half--light radii become comparable to or
smaller than the H$\alpha$ half--light radii, suggesting either that
the ionizing population dominates in the near--UV or that the
non--ionizing UV population is no more extended than the ionizing
one. Whatever the explanation, the starbursts are centrally
concentrated and are driving non--photoionization fronts located at a
minimum distance of roughly twice the extent of the starburst itself,
and a maximum (detected) distance of $\sim$15 times and $\sim$10 times
the half--light radius of the starburst, for NGC5253 and NGC3077,
respectively \citep{calz97,walt02}.

In NGC4214, both the H$\alpha$ and the U--band half--light radii are
larger than the shell's inner size, which is at variance with what
observed in the other two dwarfs. Although contamination of the
U--band half--light radius by the underlying galaxy population cannot
be excluded, the H$\alpha$ half--light radius is not affected by the
same problem, excluding measurement biases as the culprit for the
larger values. Indeed, the central starburst in this galaxy is
intrinsically far more extended than in the other two dwarfs, by about
a factor of 2, as it comprises two distinct and large HII
complexes. Conversely, the sizes of the shells are very similar to
each other in all three dwarfs. An additional difference between
NGC4214 and the other two dwarfs is in the `strenght' of the central
star formation, in the sense that NGC4214 is a less intense starburst
per unit area than the other two (Table~\ref{tbl-4}, column~7). The
star formation rate per unit area is 0.74~M$_{\odot}$~kpc$^{-2}$ in
NGC4214, a factor of about 5 and 17 lower than NGC3077
(3.7~M$_{\odot}$~kpc$^{-2}$) and NGC5253
(12.5~M$_{\odot}$~kpc$^{-2}$), respectively. For comparison, the SFR
per unit area in NGC5236 is 22.4~M$_{\odot}$~kpc$^{-2}$. The main
source of `dilution' for the NGC4214 starburst is not the overall SFR,
which is similar to the other galaxies (Table~\ref{tbl-4}, column~4),
but rather its more extended nature.

In NGC4214, our detected non--photoionization shell surrounds only one
of the two HII complexes, the northern one (Figure~\ref{fig2},
panel~e). This is similar to what found by \citet{mart98}, although
with a smaller size than our detected inner edge. If the center of
NGC4214 had contained only one HII complex, the northern one, the
starburst/shell geometries in all three dwarfs would have been almost
indistinguishable.

This fact, together with the roughly constant value of
L$_{H\alpha,nph}$/L$_{H\alpha}$, seems to suggest that external
factors, such as being an isolated or interacting galaxy, have little
or no influence on the gas ionization characteristics on the scales we
are considering. 

\subsection{Relating Non--photoionization to the Stellar 
Populations in the Starbursts}

In NGC5236, the two cavities, A and B, of non--photoionized gas may be
evolved regions within the starburst. Region~B, in particular, is
surrounded by young ($\lesssim$10~Myr, Figure~\ref{fig12}) clusters,
as identified and age--dated by \citet{harr01}. As suggested by these
authors, the starburst ringlet in NGC5236, where region~B is located,
may be experiencing an inside-out propagation, with the the youngest
stellar clusters, younger than 4--5~Myr, located further out in the
periphery of the central starburst than the 5-7~Myr old clusters. The
propagation hypothesis is further strengthened by the consideration
that region~B is more centrally located within the starburst ringlet
than the young clusters, and likely marks an earlier site of star
formation. The presence of non--photoionized gas in a localized region
removed from the main starburst and its young star cluster
component (region~C, Figure~\ref{fig12}) may be related to the dynamical
conditions of that region, rather than to the starburst itself. This
is supported by the fact that Region~C appears to be in proximity the
point of insertion of the main stellar bar onto the outer
circumnuclear ring.

Studies of the young stellar population component in NGC5253 and
NGC3077 suggest that star formation has been an on-going process for
at least 100--300~Myr in the centers of these galaxies
\citep{calz97,trem01,harr03}, thus fullfilling the requirement of long
star formation durations to provide the mechanical energy necessary to
support the observed non--photoionized gas luminosities.

In NGC5253, the UV--bright, diffuse population shows an age spread up
to $\sim$100--200~Myr \citep{calz97}, although the UV--detected
stellar clusters tend to be younger than $\sim$20~Myr
\citep{trem01,harr03}. Stars and clusters younger than $\sim$10~Myr
are concentrated in an area of $\sim$120~pc diameter around the
H$\alpha$ peak in NGC5253. The shell surrounding the starburst has a
dynamical age between $\sim$4~Myr (the closest edge to the starburst)
and 25~Myr (farthest edge), implying that the starburst has been
injecting mechanical energy into the surrounding ISM for at least that
long.

In the starburst of NGC3077, the UV--bright stellar clusters span the
age range $\sim$1--300~Myr, with clusters younger than $\sim$10~Myr
preferentially located in an area of $\sim$150~pc diameter north of
and adjacent to the CO complexes \citep{harr03}. NGC3077 appears to
display similar age--concentration characteristics as NGC5253, with
the full UV--bright population occupying a region of $\sim$470~pc in
size. However, neither here nor in the case of NGC5253, can we
discriminate whether the two galaxies are experiencing outside--in
star formation, or the older populations have had time to disperse on
larger scales than the younger ones. The CO complexes in NGC3077 are
presumably the reservoir where the most recent and the future star
formation is/will be taking place \citep{walt02}. As already remarked
in the previous section, the shells of non--photoionization
surrounding the starburst in NGC3077 have dynamical ages in the range
2--10~Myr \citep{mart98}, and therefore fit confortably within the age
span of the starburst as inferred from the stellar clusters.

The relation between the young stellar populations and the
non--photoionization regions is more difficult to infer for NGC4214,
for which the only detailed studies pertain to the most recent star
formation episode (younger than $\sim$5~Myr,
\citet{leit96,maiz98}). The shell detected in our images
(Figure~\ref{fig2}, panel~e) has dynamical age $\sim$3~Myr, and other
shells reach ages of $\sim$10~Myr \citep{mart98}. The multiwavelength
study of \citet{huch83} indicates that the on-going star formation
event started less than 100~Myr ago, but possibly more than
$\sim$20~Myr ago. As already mentioned in the previous section, the
shell of non--photoionized gas we detect is surrounding only the
northern HII complex, while we find no evidence for a similar
structure around the southern complex. This supports independent
inferences that the southern HII complex is younger than the northern
one: \citet{kobu96} find that the oxygen aboundance in southern
complex is about 0.1 dex higher than in the northen one;
\citet{maiz98} establish from measurements of H$\beta$ equivelent
widths that the southern complex is about 0.5~Myr younger than the
northern one, and may have not yet experienced supernova explosions;
\citet{hart03} find, from recent X--ray observations, evidence for an
extended hot gas component only in coincindence of the northern HII
complex, with the sourthen HII complex missing such component.

Overall, we find evidence that a significant fraction, at least 70\%
to 100\% (cf. Table~\ref{tbl-5}, column~3, and Table~\ref{tbl-6},
column~4), of the mechanical energy from starbursts is deposited into
the ISM, presumably in the form of relatively large--scale motions,
over scales of $\sim$200--1,000~pc. This provides a quantification of
the feedback mechanism from stellar winds and supernova explosions
into the galaxy's ISM. Such feedback effects have the potential to
influence future star formation in the center of the galaxies, via
self--regulation and triggering/propagation
\citep{koep95,wada99,wada01,chap01}. As importantly, the
triggering/propagation of star formation as a result of feedback
\citep{wada01,chap01} may explain the long star--forming timescales in
the centers of NGC5253 and NGC3077. Such timescale are observed to be
in the range $\approx$100-300~Myr, much longer than the expected
duration of a single starburst event (a few tens of Myr, roughly the
lifetime of massive stars). The direct triggering of the starbursts,
the interaction with the nearby galaxy(ies), happened a few hundred
Myr in the past and is not obviously active at present times. Such
long timescales for star formation can be accounted for if the
mechanical energy we observe being deposited by massive star winds and
supernova explosions is effective at triggering subsequent generations
of stars in the inhomogeneous ISM of the galaxies \citep{wada01}.

\section{Summary}

High--angular resolution imaging from HST has allowed us to
investigate the ionization nature of the gas on small-- to
intermediate--scales ($\sim$10--1000~pc) of four nearby starburst
galaxies. Our main results can be summarized as follows:
\begin{itemize}
\item Non--photoionization processes are present in all four of the
investigated galaxies. Although presence of non--photoionized gas was
already known for the three dwarfs, our images have excluded presence
of small--scale non--photoionized gas in two of them, NGC3077 and
NGC5253, and have revealed for the first time the presence of
localized non--photoionized gas in NGC5236.
\item In all cases, the projected, non--photoionized H$\alpha$
represents 3\%--4\% of the total H$\alpha$ detected in the images, but
covers up to 25\% of the H$\alpha$ emission by area in our data. Both
numbers, the fraction of non--photoionized/photoionized gas by flux
and the areal coverage fraction, should be regarded to some extent as
a lower limits. For instance, the fraction by flux increases by a
factor of a few in the metal--rich galaxies of our sample, if a less
conservative approach to separating photoionized from
non--photoionized gas is adopted. For the areal coverage fraction,
significantly larger numbers could be found by larger field
observations. Neverthless, diagnostics of distant galaxies employing
nebular emission lines (e.g., to estimate the star formation rates)
are unlikely to be significantly affected by the presence of
non--photoionization processes, even in the case of starbursts, as
such processes have a small impact on the integrated flux.
\item The current starburst can energetically sustain the observed
level of non--photoionized H$\alpha$, bringing support to the
suggestion that shocks and their precursors from massive star winds
and supernova explosions are the likely ionizing mechanism.
\item In the case of the dwarf galaxies, comparison with models show
that the energy balance argument works only if the current level of
star formation has been subtained for at least $\sim$30~Myr. Studies
of stellar populations confirm this to be the case in all three
galaxies. The basic implication is that there is strong feedback of
the starburst into the ISM, as the former is observed depositing
between 70\% and 100\% of its mechanical energy into the latter over 
scales of a few hundred pc. 
\item No such requirement is needed for the giant spiral, where an
ongoing starburst triggered $\sim$10~Myr ago can still provide enough
energy input to account for the observed level of
non--photoionization.
\item In all four cases, regions of non--photoionzation and CO regions do
not coincide spatially. This is further support to the idea that
molecular regions host the youngest or future star formation, which
has not experienced supernova explosions yet.
\item In the three dwarfs non--photoionization is located mainly in
large--scale shells, while in NGC5236 it is located in small--scale
cavities surrounded by recent star formation. The cavities may mark
sites of earlier star formation within the current starburst. No
large--scale ($\sim$200-300~pc) non--photoionization has been
identified in the large spiral. Thus, star formation remains a local
event, confined by the deep potential well of this massive galaxy.
\item The shell/starburst geometries, the shell 
morphologies, and the non--photoionized/photoionized gas flux ratios are 
similar among the three dwarf galaxies, suggesting that external factors, 
such as being isolated or interacting galaxies, have negligible influence 
on the gas ionization conditions on the $\sim$few$\times$100~pc scale 
considered.  
\item In the images of NGC4214 a candidate supernova remnant, removed
from the main sites of recent star formation, has been located. This
candidate SNR is distinct from the SNR identified by \citet{mack00}.
\end{itemize}

\acknowledgments

This work has been supported by the NASA LTSA grant NAG5--9173 and by
the NASA HST grant GO--9144. D.C. thanks Nino Panagia for stimulating
discussions on line emission characteristics in photoioned regions and
in supernova remnants, and Fabian Walter for providing his CO
images of NGC3077. The authors would like to thank the anonymous 
referee for the many suggestions that have helped improve the 
manuscript.



\clearpage


\begin{figure}
\caption{Line and continuum intensity and ratio images are shown for
NGC3077, where each bin is 3$\times$3~pixels (or
0.3$^{\prime\prime}\times$0.3$^{\prime\prime}$) and the total size of
each image is 74$^{\prime\prime}$.5: (a) UV image; (b) H$\alpha$
emission; (c) [SII]/H$\alpha$; (d) [OIII]/H$\beta$; (e) Locus of the
non--photoionized areas, overplotted in black onto the H$\alpha$
image. The North-East direction is indicated by the vector. Light
areas represent low values, dark areas are high values of intensity or
ratio.  In panel~(e), the white ellipses show the approximate position
of the CO detections of \citet{walt02}. The white broken polygons mark
the approximate position of the shells identified in \citet{mart98},
using the same naming convention.
\label{fig1}}
\end{figure}

\clearpage 
\begin{figure}
\caption{As Figure~\ref{fig1}, for NGC4214. The total size of each
image is 111$^{\prime\prime}$. In panel~(e), a number of artifacts are
present that appear as non--photoionization regions: some of the
background in the PC chip and the seams between the four WFPC2
chips. These have been excluded in all discussions and
calculations. The thick white polygons show the position of the
H$\alpha$ cavity (As) and of the candidate supernova remnant
(SNR). The large thin polygon (A) marks the approximate position of
the central shell identified in \citet{mart98}, using the same naming
convention. The white circles show the approximate position of the CO
detection peaks of \citet{beck95}.\label{fig2}}
\end{figure}

\clearpage 
\begin{figure}
\caption{As Figure~\ref{fig1}, for NGC5236. In this case, bins of
3$\times$3~pixels correspond to
0.138$^{\prime\prime}\times$0.138$^{\prime\prime}$. The total size of
each image is 34$^{\prime\prime}$.5. In panel~(e), the 
non--photoionized regions
discussed in the text are marked by polygons and named `A', `B', and
`C'; region `C' is the non--photoionized region located close to the
insertion point of the main stellar bar onto the outer circumnuclear
ring.  A few other regions of non--photoionized gas are in reality
cases of oversubtracted underlying continuum in the H$\alpha$
image. The black circles identify the approximate position of the
peaks of CO emission in the center of the galaxy
\citep{sofu94,isra01}. 
\label{fig3}}
\end{figure}

\clearpage 
\begin{figure}
\caption{As Figure~\ref{fig1}, for NGC5253. The total size of each
image is 71$^{\prime\prime}$. In panel~(e), the white ellipse shows
the approximate location of the CO detected by \citet{turn97} and by
\citet{meie02}.  The CO is mainly located along the dust lane in this
galaxy \citep{meie02}, with other clouds located outside the
field-of-view shown here. \label{fig4}}
\end{figure}

\clearpage 

\begin{figure}
\plottwo{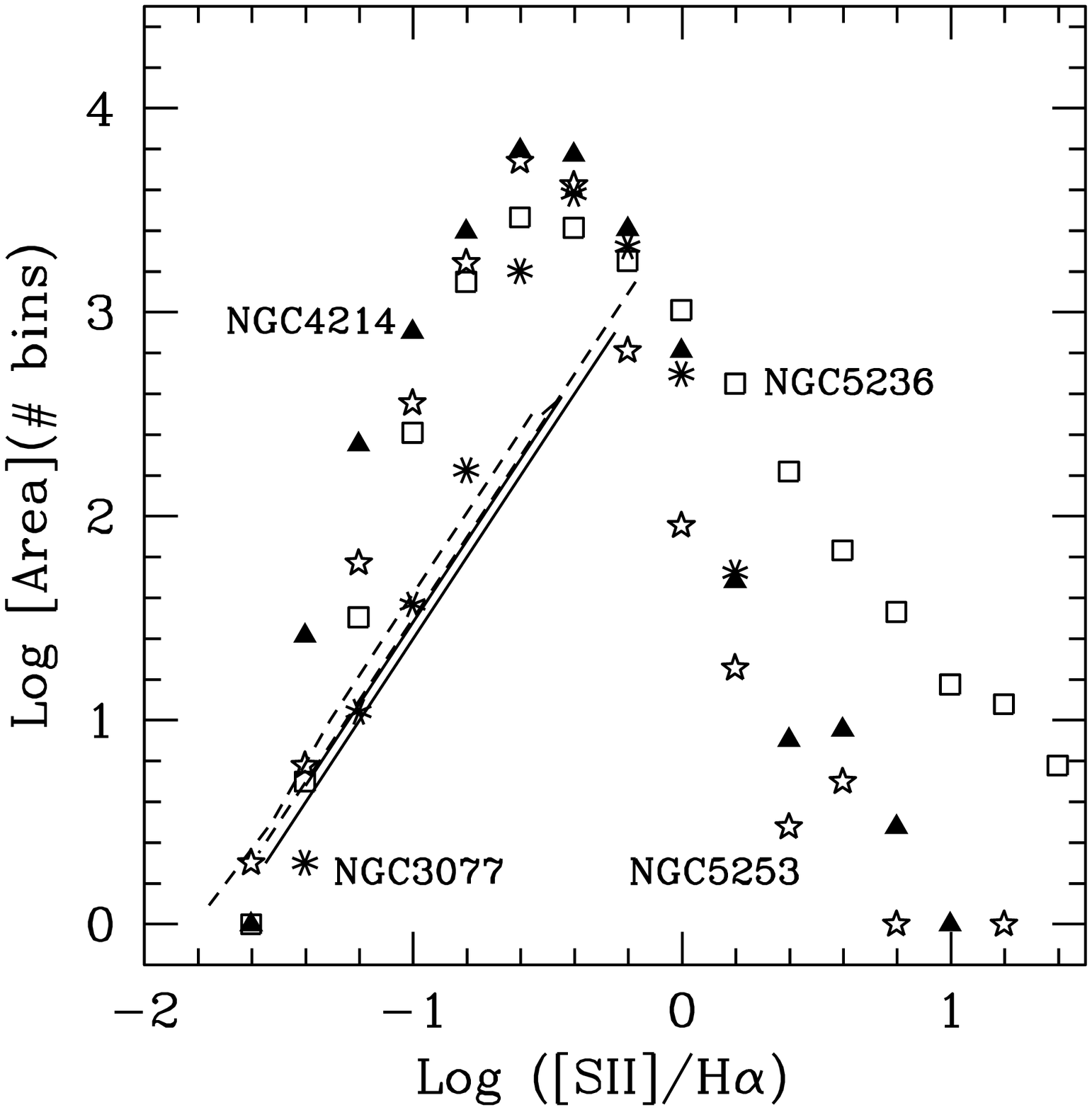}{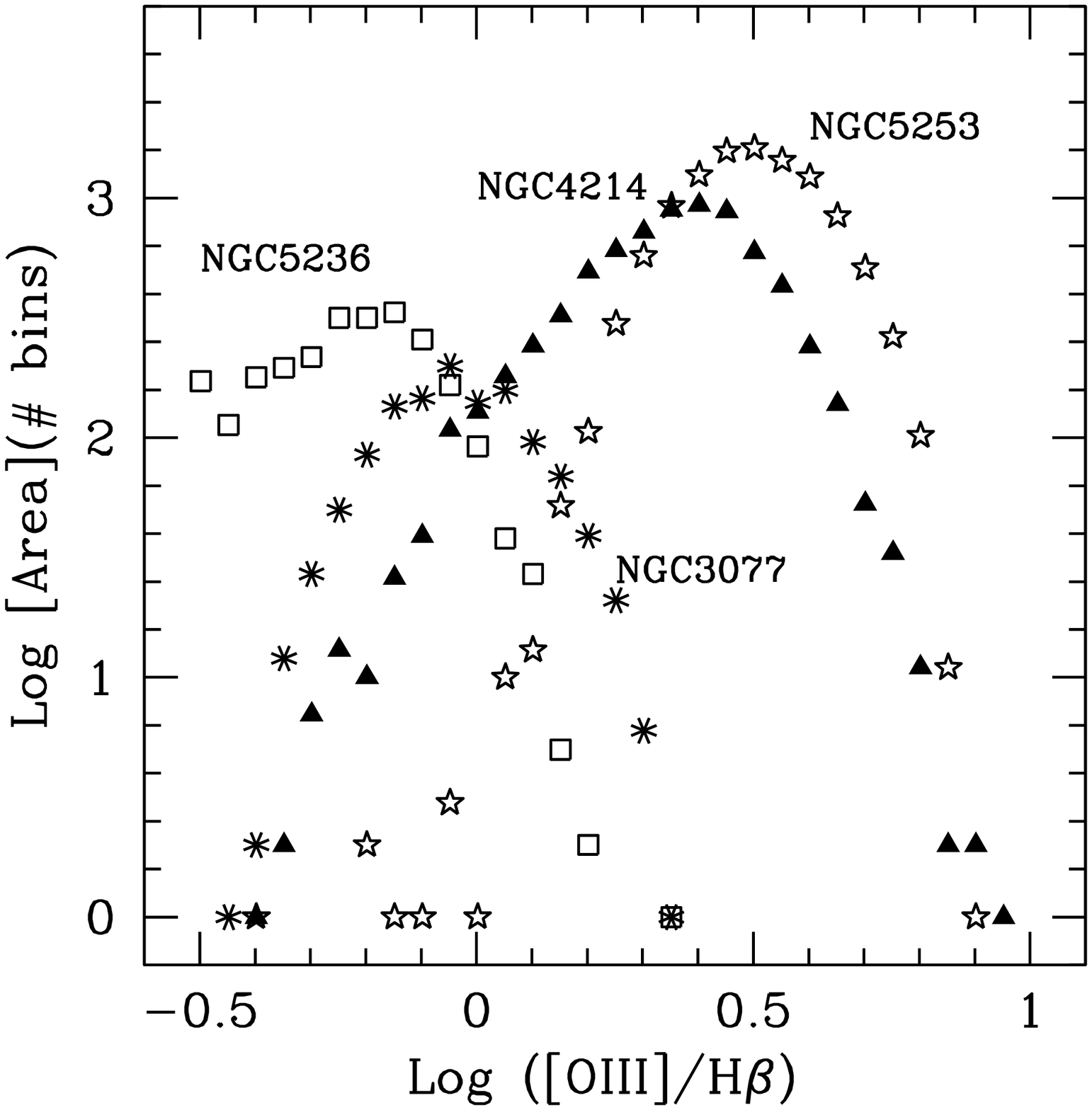}
\caption{Histograms of the [SII]/H$\alpha$ (left) and [OIII]/H$\beta$ 
(right) ratios, plotted as the number of 3$\times$3~pixel bins occupied by
different values of the ratio. The symbols represent: NGC3077
(asterisks), NGC4214 (filled triangles), NGC5236 (squares), NGC5253
(stars). The continuous and dashed lines in the right-hand-side panel
show the locus of [SII]/H$\alpha$ line ratio for gas photoionized by a
central point-like source; continuous lines from the Cloudy models as
described in \citet{mart97}, dashed lines from the models of \citet{kewl01}. 
See text for more details.\label{fig5}}
\end{figure}

\clearpage 

\begin{figure}
\plotone{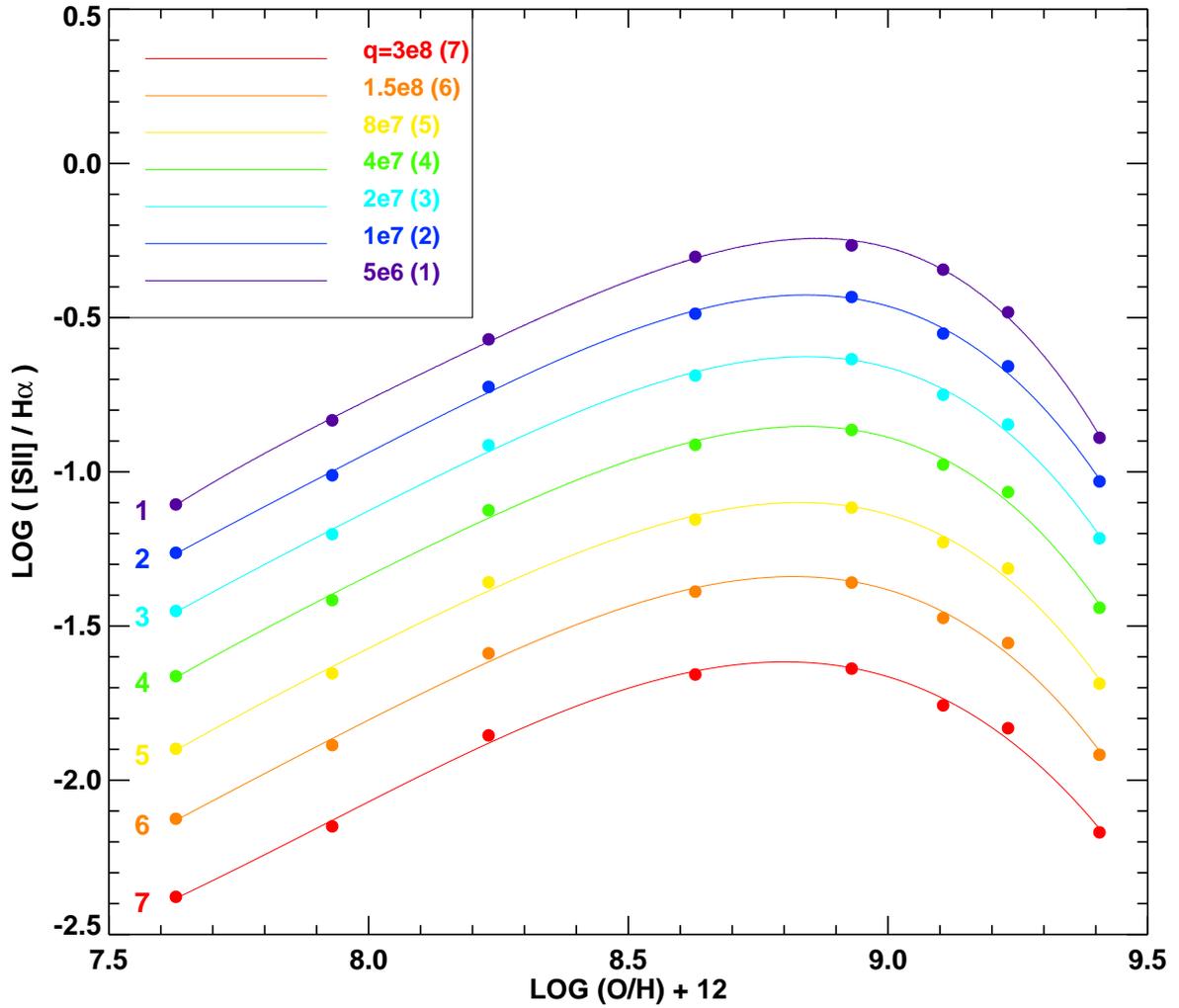}
\caption{The [SII]/H$\alpha$ ratio as a function of oxygen abundance
for a range of ionization parameters q (where q is related to the
commonly used ionization parameter U=q/c), from the models of
\citet{kewl01}. The paramater range shown brackets the expected
conditions in starburst galaxies.\label{fig6}}
\end{figure}

\clearpage 

\begin{figure}
\caption{Emission line diagnostic diagram for the four galaxies. For
each galaxy, the datapoints represent the 3$\times$3~pixel bins above
a 5~$\sigma$ sensitivity cut in common between the two ratio images.
Photoionization models at the appropriate metallicity (symbols on
lines, and labels) and the `maximum starburst line' (continuous line)
from \citet{kewl01} are shown, together with the position of
shock--excited emission line ratios (indicated by the letter `S') from
\citet{shul79}. Median 1~$\sigma$ uncertainty bars are shown at the
right--bottom corner of each panel; the random uncertainties for the
[OIII]/H$\beta$ ratios have been convolved with the uncertainty for
the CTE losses. The photoionization models span a range of parameters;
each photoionization line is identified by a metallicity value (Z0.2,
Z0.5, etc., referred to solar metallicity) and by an electron density
value (n=10 or 350~cm$^{-3}$). The range of values of the ionization
parameter q=0.5,1.0,2.0,4.0,8.0,15,30$\times$10$^7$~cm~s$^{-1}$ is
shown by symbols on each photoionization line. The ionizing population
is given by continuous star formation, although for the Z2.0 case
instantaneous burst models are also considered (labelled by
`i'). Shock models are shown for a range of parameters: S$_D$,
S$_E$, and S$_G$ correspond to shocks with cosmic abundance, gas
density of 10~cm$^{-3}$, and velocities $v=$90, 100, and
130~km~s$^{-1}$, respectively; S$_J$ corresponds to shocks with
depleted abundances and $v=$100~km~s$^{-1}$. The total number of bins
available for each galaxy in this plot is determined by the shallower
[OIII]/H$\beta$ ratio images (see text). For NGC5253, the locus of the
data points from ground--based images (from \citet{calz99}) is shown
as a rectangular region. The dotted straight lines in the four panels
mark the adopted boundary between photonized and non--photoionized
data points for each galaxy. The data points to the top--right of
these lines, combined with the data points above the horizontal lines
of Figure~\ref{fig10}, are used to derive the `non--photoionization
images' (panels e) in Figures~\ref{fig1}--\ref{fig4}.
\label{fig7}}
\end{figure}

\clearpage 

\begin{figure}
\caption{The same as Figure~\ref{fig7} for NGC3077 and NGC4214, where the 
sensitivity cut has been placed to 3~$\sigma$. More data points in the non--photoionized region (right and above the `maximum starburst line') are present 
in this case than in Figure~\ref{fig7}.\label{fig8}}
\end{figure}

\clearpage 

\begin{figure}
\caption{The same as Figure~\ref{fig7} for NGC5236. In the bottom
panel, the H$\alpha$ images have been corrected for the [NII]
contribution assuming that the ratio [NII]/H$\alpha$ is constant
(section~3.1.2); the top panel reports for comparison the same plot as
Figure~\ref{fig7}, where the H$\alpha$ image has been corrected for
the [NII] contribution assuming [NII]$\propto$[SII].
\label{fig9}}
\end{figure}

\clearpage 

\begin{figure}
\caption{Line ratio [SII]/H$\alpha$ as a function of the surface
brightness H$\alpha$, $\Sigma_{H\alpha}$. The surface brightness is
normalized to the half-light radius surface brightness
$<\Sigma>$. High-valued [SII]/H$\alpha$ bins are preferentially
located in areas of low H$\alpha$ surface brightness. Median
1~$\sigma$ uncertainty bars are shown at the left corners of each
panel. The sharp cut-off to the right of each diagram is the
5~$\sigma$ sensitivity limit to both the line ratios and the line
intensity images. The horizontal line in each panel gives a
representative positioning of the locus above which the gas is ionized
by non-radiative processes (from the `maximum starburst line' of
Figure~\ref{fig7}, \citet{kewl01}). For NGC5236, the split of points
in two distinct loci along the H$\alpha$ surface brightness axis is
due to the availability of reddening corrections for only a fraction
of the datapoints (those located in the right-hand-side locus); this
is due to the H$\beta$ image being shallower than the H$\alpha$ and
[SII] images. The same is true for all four galaxies, but the effect
is most obvious in NGC5236 because of its more extreme reddening
corrections (see Table~\ref{tbl-1}).\label{fig10}}
\end{figure}

\clearpage 

\begin{figure}
\caption{Line ratio [OIII]/H$\beta$ as a function of the surface
brightness H$\alpha$, $\Sigma_{H\alpha}$. As in Figure~\ref{fig10},
the surface brightness is normalized to the half-light radius surface
brightness $<\Sigma>$. Median 1~$\sigma$ uncertainty bars are shown at
the left corners corner of each panel; the random uncertainties for
the [OIII]/H$\beta$ ratios have been convolved with the uncertainty
for the CTE losses. The images clearly show that the [OIII]/H$\beta$
ratios reach shallower limits in H$\alpha$ surface brightness than the
[SII]/H$\alpha$ ratios (cf. Figure~\ref{fig10}).\label{fig11}}
\end{figure}

\clearpage 

\begin{figure}
\caption{The age map of the clusters in the center of NGC5236 (from
Figure~11 of \citet{harr01}) is shown with the outlines of the
non--photoionized regions A, B, and C superimposed. North is up, East
is left. The figure shows that regions~A and B are embedded in the
area of recent star and cluster formation, while region~C is more at
the periphery of that area. Because of its location at the insertion
point of the main stellar bar on the outer circumnuclear ring, 
region~C may be unrelated to the starburst proper.\label{fig12}}
\end{figure}






\clearpage

\begin{deluxetable}{lrrrrrrrrrrr}
\tabletypesize{\scriptsize}
\tablecaption{Characteristics of the Program Galaxies. \label{tbl-1}}
\tablewidth{0pt}
\tablehead{
\colhead{Galaxy} & \colhead{Morph.\tablenotemark{a}} & \colhead{v$_H$} 
& \colhead{D\tablenotemark{b}}  & \colhead{d\tablenotemark{c}} 
& \colhead{S\tablenotemark{d}} & \colhead{s\tablenotemark{e}}
& \colhead{M$_B$\tablenotemark{f}} 
& \colhead{E(B$-$V)$_G$\tablenotemark{g}} 
& \colhead{E(B$-$V)$_I$\tablenotemark{h}}
& \colhead{Log(L$_{IR}$/L$_B$)\tablenotemark{i}} 
& \colhead{(O/H)\tablenotemark{j}}  \\
\colhead{Name} & \colhead{Type} & \colhead{(km/s)} & \colhead{(Mpc)} 
& \colhead{(kpc)} & \colhead{(kpc)} & \colhead{(pc)} & \colhead{(mag)} 
& \colhead{(mag)} & \colhead{(mag)} & \colhead{} & \colhead{} 
}
\startdata
NGC~3077 &I0 pec & 14&3.85$\pm$0.3 &6.1 &1.4 &5.6&$-$17.5 &0.07 &0.56 &$-$0.486 &8.9\\
NGC~4214 &IAB(s)m&291&2.94$\pm$0.18&7.3 &2.0 &4.3&$-$17.2 &0.02 &$\sim$0.15  &$-$0.466 &8.22\\
NGC~5236 &SAB(s)c&516&4.5$\pm$0.8  &16.9&0.75 &3.0&$-$20.3 &0.06 &0.28  &$-$0.148 &9.17\\
NGC~5253 &Im pec &404&4.0$\pm$0.3  &5.8 &1.5 &5.8&$-$17.5 &0.06 &0.01 &$-$0.238 &8.23\\
 \enddata


\tablenotetext{a}{Morphological types from the RC3 \citep{deva91}.}
\tablenotetext{b}{Adopted distances, from \citet{saka01} (NGC3077),
\citet{maiz02} (NGC4214), and \citet{thim03} (NGC5236 and NGC5253).}
\tablenotetext{c}{Physical size of the galaxy, along the major axis; 
the major axis is the B=25~mag isophotal diameter from RC3.}
\tablenotetext{d}{Physical size subtended by the relevant WFPC2
chip(s): WF3 for NGC3077 and NGC5253, PC for NGC5236, and the entire
WFPC2 FOV for NGC4214.}  
\tablenotetext{e}{Physical size subtended by
three WFPC2 pixels, using the angular scale of the relevant chip for
each galaxy: WF for NGC3077, NGC4214, and NGC5253, and PC for
NGC5236.}  
\tablenotetext{f}{Absolute B magnitude calculated from B$_T^0$.}  
\tablenotetext{g}{Galactic extinction from \citet{schl98},
as reported in NED.}  
\tablenotetext{h}{Mean internal extinction of
the central starburst region, as derived by \citet{calz94} and 
\citet{stor94} for the inner $\sim$15$^{\prime\prime}$ of NGC3077, 
NGC5253, and NGC5236, and by \citet{maiz98} for the inner 
$\sim$50$^{\prime\prime}$ of NGC4214.}  
\tablenotetext{i}{The infrared-to-blue ratio. L$_{IR}$ is
derived from the IRAS fluxes at 60~$\mu m$ and 100~$\mu m$ using the
formula of \citet{helo88}. L$_B$ is the B-band luminosity defined as
$\lambda$f($\lambda$); the flux density f($\lambda$) is from B$_T^0$
corrected for Galactic extinction (see column~8 of this Table).}
\tablenotetext{j}{Oxygen abundances, 12$+$log(O/H), from
\citet{mart97} (NGC4214 and NGC5253), and \citet{zari94} (NGC5236).
For NGC3077 we used the strong emission lines from the spectrum of
\citet{mcqu95} together with the models of \citet{kewl02} to derive an
oxygen abundance of 8.9$\pm$0.1.}  
\tablecomments{Quantities for which a reference is not given are from
the NASA Extragalactic Database (NED).\\
NGC3077 is a member of the M81 cluster, in close
interaction with M81 itself and M82. NGC4214 is an isolated
dwarf. NGC5236 and NGC5253 are likely an interacting pair.}

\end{deluxetable}


\clearpage

\begin{table}
\caption{Summary of the HST/WFPC2 Observations.\label{tbl-2}}
\scriptsize
\begin{tabular}{llrrrrrrrr}
\tableline\tableline
Filter & Band & \multicolumn{2}{c}{NGC3077\tablenotemark{a}} & \multicolumn{2}{c}{NGC4214\tablenotemark{a}}
& \multicolumn{2}{c}{NGC5236\tablenotemark{a}} & \multicolumn{2}{c}{NGC5253\tablenotemark{a}} \\
       &     & Prog ID & Exp. time (s) & Prog ID & Exp. time (s) 
& Prog ID & Exp. time (s) & Prog ID & Exp. time (s) \\
\tableline
F255W & UV      &      &               &         &              
&      &                  & 6124    & 3x700,6x800\\
F300W & UV      & 9144 & 3x800   &    &      & 8234    & 3x700   
&  &  \\
F336W &$\sim$UV &      &               & 6569  & 260,2x900 
&  &  &  & \\
F487N &H$\beta$ & 9144 & 3x700,1300    & 9144 & 2x600,1000 
& 8234 & 1000,1100,1200 & 6524 & 1200,3x1300\\
F502N &[OIII] & ''   & 350,600,800 & 6569 & 700,800  
& ''   & 2x1200  & 9144 & 200,260,600,800\\
F547M &$\sim$V              & ''   & 2x600           &      &             
& ''   & 180,350,400 & 6524 & 2x200,2x600\\
F555W &$\sim$V              &      &                 & 6569 & 100,2x600 
&      &                    &      &       \\
F656N &H$\alpha+$[NII] & 9144 & 300,2x800     &  ''  & 2x800       
& 8234 & 2x600            & 6524 & 2x500,1100,1500 \\
F673N &[SII] & ''& 2x700,1400    & 9144 & 2x600,1100 
& ''   & 2x1200          & 9144 & 2x600,1200\\
F814W &$\sim$I              &  ''  & 300,400       & 6569 & 100,2x600 
& ''   & 160,200,350         & 6524   & 2x180,2x400\\
\tableline
\end{tabular}

\tablenotetext{a}{Exposure times are given for individual exposures in each
filter/galaxy combination. Observations in each band consist of 2--4
individual exposures. Short exposures in some sets (e.g., 100~s exposures in
F555W and F814W of NGC4214) were obtained with the purpose of correcting
longer exposures for the effects of saturation, and were not combined with the
deeper images.}

\tablecomments{The restframe wavelengths of the emission lines are:
H$\beta ~\lambda$4861~\AA, [OIII]~$\lambda$5007~\AA, H$\alpha
~\lambda$6563~\AA~ $+$ [NII]~$\lambda$6548~\AA, and
[SII]~$\lambda\lambda$6726,6731~\AA. The stronger of the two [NII] 
lines, [NII]~$\lambda$6584~\AA, falls within the F656N bandpass only in the
case of NGC3077, the galaxy with the smallest redshift.}
\end{table}

\clearpage

\begin{table}
\caption{Corrections to and Sensitivity Limits of Emission Line Images.\label{tbl-3}}
\scriptsize
\begin{tabular}{lrrrrrrrr}
\tableline\tableline
Band & \multicolumn{2}{c}{NGC3077} & \multicolumn{2}{c}{NGC4214}
& \multicolumn{2}{c}{NGC5236} & \multicolumn{2}{c}{NGC5253} \\
     & T$_{corr}$\tablenotemark{a} & 1~$\sigma$ Limit\tablenotemark{b}
& T$_{corr}$\tablenotemark{a} & 1~$\sigma$ Limit\tablenotemark{b}
& T$_{corr}$\tablenotemark{a} & 1~$\sigma$ Limit\tablenotemark{b}
& T$_{corr}$\tablenotemark{a} & 1~$\sigma$ Limit\tablenotemark{b} \\
\tableline
H$\beta$ &$<$0.2\% &1.34E$-$17 & 1\% &1.62E$-$17 &1.8\% &1.40E$-$17 &1.3\% &1.03E$-$17\\ 
$[OIII]$   & 1\%  &1.95E$-$17 &$<$0.2\% &1.62E$-$17 &10\% &1.35E$-$17 &1.8\% &2.21E$-$17\\
H$\alpha$& 1\% &9.14E$-$18&7.4\% &7.60E$-$18 &22.5\% &9.39E$-$18 &9.3\% &6.75E$-$18\\
([NII]-corr.) & $-$3.7\%  &   & $-$2.4\% & & $-$16\% & & $-$1.4\% & \\  
$[SII]$    & 2\%     &6.86E$-$18 &0.5\% &5.91E$-$18 & 1\% &7.11E$-$18 & 0.6\% &7.29E$-$18\\
\tableline
\end{tabular}

\tablenotetext{a}{Filter transmission curve correction, in percent, 
to the emission line flux. The transmission curves for the filters were
obtained from the package STSDAS/SYNPHOT. The sense of the correction is to
increase the line flux. For H$\alpha$, the correction in
percent due to [NII] contamination is indicated in the second line, and this
correction decreases the line flux.}
\tablenotetext{b}{Emission line 1~$\sigma$ detection limit, in units of 
erg~s$^{-1}$~cm$^{-2}$~bin$^{-1}$, where a bin=3x3 pix$^2$, after subtraction 
of the stellar continuum.}

\end{table}

\clearpage

\begin{table}
\caption{Measured and Derived Quantities.\label{tbl-4}}
\scriptsize
\begin{tabular}{lcccccccc}
\tableline\tableline
Galaxy & F$_{H\alpha}$\tablenotemark{a} & L$_{H\alpha}$\tablenotemark{b} 
& SFR$_{H\alpha}\tablenotemark{c}$ 
& Q\tablenotemark{d} & R$_S$\tablenotemark{e} & $<\Sigma>_{H\alpha}$\tablenotemark{f}& R$_{\Sigma}$\tablenotemark{g} & R$_{NUV}$\tablenotemark{h}\\
Name & (erg~s$^{-1}$~cm$^{-2}$) & (erg~s$^{-1}$) & (M$_{\odot}$~yr$^{-1}$) 
& (ph~s$^{-1}$) & (pc) 
& (erg~s$^{-1}$~cm$^{-2}$~arcsec$^{-2}$) & (pc) & (pc)\\
\tableline
NGC3077 & 5.41E$-$12 & 9.6E$+$39 & 0.076 & 7.1E$+$51 & 61, 111 & 2.95E-14 & 101 & 129\\
NGC4214 & 1.05E$-$11 & 1.1E$+$40 & 0.089 & 8.0E$+$51 & 79, 115 & 5.72E-15 & 245 & 282\\
NGC5236 & 1.60E$-$11 & 3.9E$+$40 & 0.308 & 2.9E$+$52 & 121, 177& 1.76E-13 & 83 & 96\\
NGC5253 & 1.75E$-$11 & 3.4E$+$40 & 0.270 & 2.5E$+$52 & 116, 168& 9.55E-14 & 104 & 157\\
\tableline
\end{tabular}

\tablenotetext{a}{H$\alpha$ flux in the WFPC2 chip(s) for each galaxy, 
corrected pixel-by-pixel for underlying 
stellar absorption and Galactic and internal dust attenuation.}
\tablenotetext{b}{H$\alpha$ luminosity, derived from the fluxes in the 
previous column.}  
\tablenotetext{c}{Star formation rate, from the H$\alpha$ luminosity and 
the conversion formula of \citet{kenn98}.}  
\tablenotetext{d}{Number of ionizing photons, assuming constant star
formation \citep{leit99}.}  
\tablenotetext{e}{Stromg\"en radii 
R$_S$=(3Q/4$\pi\alpha_B$n$^2 \epsilon$)$^{(1/3)}$. 
The left-hand-side value is calculated
for a gas density n=100~cm$^{-3}$ and a filling factor
$\epsilon$=0.05; the right-hand-side value for n=40~cm$^{-3}$ and
$\epsilon$=0.1 \citep{mart97}. In both cases,
$\alpha_B$=2.6$\times$10$^{-13}$~cm$^3$~s$^{-1}$ \citep{oste89}. }
\tablenotetext{f}{H$\alpha$ surface brightness, calculated from the total H$\alpha$ flux divided by the area of the half--light radius.}
\tablenotetext{g}{H$\alpha$ half--light radius.}
\tablenotetext{h}{Near--UV half--light radius. The near--UV
information comes at different wavelengths for different galaxies
(Table~\ref{tbl-2}).  Some additional caveats: insufficient dust
corrections (expected to be important in NGC3077, NGC5236, and
NGC5253), which tend to be larger towards the centers of the galaxies,
and contributions from a population underlying the starburst (expected
to affect NGC4214, which is measured in a U filter, rather than a near--UV
filter) may artificially cause the UV radii to be larger than
the actual starburst's size.}

\end{table}

\clearpage

\begin{table}
\rotate
\caption{Measured Characteristics of the Non--Photoionized Gas.\label{tbl-5}}
\scriptsize
\begin{tabular}{lrrllcl}
\tableline\tableline
Galaxy & F$_{H\alpha,nph}$\tablenotemark{a} 
& L$_{H\alpha,nph}$\tablenotemark{b} 
& L$_{H\alpha,nph}$/L$_{H\alpha}$\tablenotemark{c}  & A$_{nph}$/A$_{tot}\tablenotemark{d}$
& R$_{shell}$\tablenotemark{e} & R$_{cavity}$\tablenotemark{f}\\
Name & (erg~s$^{-1}$~cm$^{-2}$) & (erg~s$^{-1}$) &  &   & (pc)  & (pc)\\
\tableline
NGC3077 & 2.33E$-$13 & 4.1E$+$38 & 0.043 & 0.24 & 179 &  ... \\
NGC4214 & 4.31E$-$13 & 4.5E$+$38 & 0.041 & 0.21 & 215  & 35, 43 \\
NGC5236 & 4.82(1.92)E$-$13 & 1.2(0.5)E$+$39 & 0.030(0.012) & 0.16(0.07) & ... & 56, 21 (45, 13)\\
NGC5253 & 5.60E$-$13 & 1.1E$+$39 & 0.032 & 0.18 & 221 & ... \\
\tableline
\end{tabular}

\tablenotetext{a}{Flux in H$\alpha$ of the gas not ionized by
radiative processes, as derived from the 3~$\sigma$ line ratio images. 
For NGC5236,  the second number (in
parenthesis) is calculated from H$\alpha$ images that have been corrected for 
the [NII] contribution using the assumption [NII]$\propto$H$\alpha$, rather 
than the default[NII]$\propto$[SII] (see discussion in section~5.1).}
\tablenotetext{b}{Luminosity in H$\alpha$ of the non--photoionized gas. 
For the numbers in parenthesis, see note (a).}  
\tablenotetext{c}{Fraction of the total H$\alpha$ luminosity associated with 
non--photoionization processes. 
For the numbers in parenthesis, see note (a).}  
\tablenotetext{d}{Fraction of area occupied by the non--photoionized gas. 
The two areas, A$_{nph}$ and A$_{tot}$, are determined from bins of 
positive detection in the line ratio images. For the numbers in parenthesis, 
see note (a).}
\tablenotetext{e}{Characteristic radius of the non--photoionized
regions with shell and/or filamentary morphology surrounding the central
starburst. The listed figures refer to the inner edge of the `shells', as the
size of the outer edge is dictated by the sensitivity limit of our images. In
the case of NGC4214, the `shell' of diffuse non--photoionized 
gas appears to trace the edges of Region~A of \citet{mart98}.}
\tablenotetext{f}{Characteristic radius of the non--photoionized regions with
morphology of `cavities' or enclosed regions. In the case of NGC4214,
the first number refers to region~As; the second number is the
size of the candidate supernova remnant in the PC chip. In the case of
NGC5236, the first number refers region~A and the second to region~B. 
For the numbers in parenthesis, see note (a).}

\end{table}

\clearpage

\begin{table}
\caption{Predictions for Shocked Gas.\label{tbl-6}}
\scriptsize
\begin{tabular}{lccccccc}
\tableline\tableline
Galaxy &  L$_{H\alpha}$\tablenotemark{a} & L$_{mech}$\tablenotemark{b} 
& L$_{H\alpha, mech}$/L$_{H\alpha}$\tablenotemark{c} & F$_{NUV}$\tablenotemark{d} 
& L$_{NUV}$\tablenotemark{e} &  L$_{mech}$\tablenotemark{f} 
& L$_{H\alpha, mech}$/L$_{H\alpha}$\tablenotemark{g} \\
Name & (erg~s$^{-1}$) & (erg~s$^{-1}$) & 
& (erg~s$^{-1}$~cm$^{-2}$~\AA$^{-1}$) & (erg~s$^{-1}$~\AA$^{-1}$) 
& (erg~s$^{-1}$) & \\
\tableline
NGC3077 & 9.6E$+$39 & (0.85--2.3)E$+$40  
&0.022--0.060 & 8.7E$-$14 & 1.5E$+$38 & (1.4--2.3)E$+$40 
& 0.038--0.058\\
NGC4214 & 1.1E$+$40 & (0.51--1.67)E$+$40 
& 0.012--0.038 & 1.4E$-$13 & 1.5E$+$38 & (1.1--2.2)E$+$40 
& 0.025--0.051\\
NGC5236 & 3.9E$+$40 & (6.1--11.7)E$+$40 
& 0.038--0.074 & 2.9E$-$13 & 7.1E$+$38 & (9.9--13.6)E$+$40 
& 0.061--0.087\\
NGC5253 & 3.4E$+$40 & (1.6--5.1)E$+$40 
& 0.012--0.038 & 2.5E$-$13 & 4.9E$+$38 & (2.2--4.5)E$+$40 
& 0.016--0.032\\
\tableline
\end{tabular}

\tablenotetext{a}{H$\alpha$ luminosity within the relevant WFPC2 chip(s), 
reported from column~3 of Table~\ref{tbl-4}.}
\tablenotetext{b}{Mechanical luminosity as derived from the 
L$_{H\alpha}$ due to photoionization (previous column) and the
models of \citet{leit99}. The range in values of L$_{mech}$
corresponds to the age range 10$^7$--10$^8$~yr of the stellar
populations, adopting a Salpeter IMF to 100~M$_{\odot}$; for each
galaxy, the available models with the closest metallicity value are
considered. For continuous star formation
models the ratio of mechanical luminosity to number of ionizing
photons, L$_{mech}$/N$^o_{ion}$ (and thus L$_{mech}$/L$_{H\alpha}$), 
levels off and become constant after$\sim$3$\times$10$^7$~yr.}
\tablenotetext{c}{The predicted fraction of H$\alpha$ luminosity due to
non--photoionization processes. L$_{H\alpha, mech}$ is derived from the 
photoionized H$\alpha$ flux, adopting a 2.5\% fraction of the mechanical
energy radiated in the H$\alpha$ line \citep{bine85}. These 
numbers should be compared with the measurements of column~4 in 
Table~\ref{tbl-5}.}
\tablenotetext{d}{Near--UV flux in the relevant WFPC2 chip(s) for each 
galaxy, corrected for Galactic foreground
extinction and pixel--by--pixel internal dust attenuation (for the
latter, using the formula of \citet{calz01}). None of the F$_{NUV}$ values 
has been corrected for total absorption, i.e., the fraction of the near--UV 
fluxes completely absorbed by dust. The near--UV passband is different for
different galaxies (see Table~\ref{tbl-2}): 
F255W has pivot wavelength 2600~\AA, F300W  has pivot wavelength 2990~\AA, 
and F336W has pivot wavelength 3340~\AA~ \citep{bire02}.}
\tablenotetext{e}{Near--UV luminosity, derived
from the fluxes in the previous column.}  
\tablenotetext{f}{Mechanical luminosity, derived from the near--UV
luminosity and the same stellar population model from Starburst99
\citep{leit99} used for the H$\alpha$ luminosity above.}
\tablenotetext{g}{The predicted fraction of H$\alpha$ luminosity due to
non--photoionization processes. L$_{H\alpha, mech}$ is derived from the 
number in the previous column. As before, we adopt a 2.5\% fraction of the 
mechanical energy radiated in the H$\alpha$ line \citep{bine85}. These 
numbers should be compared with the predictions of column~4 in this Table and 
the measurements of column~4 in Table~\ref{tbl-5}.}

\end{table}



\end{document}